\documentstyle[12pt,aasms4]{article}
\def\sur#1#2{{#1 \over #2}}
\def\lta{\mathrel{\spose{\lower 3pt\hbox{$\mathchar"218$}}
     \raise 2.0pt\hbox{$\mathchar"13C$}}}
\def\gta{\mathrel{\spose{\lower 3pt\hbox{$\mathchar"218$}}
     \raise 2.0pt\hbox{$\mathchar"13E$}}}
\def\spose#1{\hbox to 0pt{#1\hss}}

\def\etal{et al. }

\slugcomment{Accepted for publication in ApJ}
\lefthead{Balland, Silk \& Schaeffer}
\righthead{Collision Induced Galaxy Formation}
\begin{document}
\title{COLLISION INDUCED GALAXY FORMATION}
\author{Christophe Balland\altaffilmark{1}
\altaffiltext{1}
{Also Observatoire Astronomique de
Strasbourg and Universit\'e d'Aix-Marseille II, France}
}
\and
\author{Joseph Silk}
\affil{University of California\\
Center for Particle Astrophysics\\
and Astronomy Department\\
Berkeley, USA\\}
\and
\author{Richard Schaeffer}
\affil{C.E.N., Saclay, FRANCE}

\begin{abstract}

We present a semi-analytical model in which galaxy collisions and
strong tidal interactions, both in the field and during the collapse
phase of groups and clusters help determine galaxy morphology.  From a
semi-analytical analysis based on simulation results of tidal
collisions (Aguilar \& White 1985), we propose simple rules for energy
exchanges during collisions that allow to discriminate between
different Hubble types: efficient collisions result in the disruption
of disks and substantial star formation, leading to the formation of
elliptical galaxies; inefficient collisions allow a large gas
reservoir to survive and form disks.  Assuming that galaxy formation
proceeds in a $\Omega_0=1$ Cold Dark Matter universe, the model both
reproduces a number of observations and makes predictions, among which
are the redshifts of formation of the different Hubble types in the
field. When the model is normalized to the present day abundance of
X-ray clusters, the amount of energy exchange needed to produce
elliptical galaxies in the field implies that they formed at $z\gta
2.5$ while spiral galaxies formed at $z\lta 1.5$. The model also
offers a natural explanation for biasing between the various
morphological types.  We find that the present day morphology-density
relation in the field is well reproduced under the collision
hypothesis. Finally, predictions of the evolution of the various
galaxy populations with redshift are made, in the field as well as in
clusters.

\end{abstract}

\keywords{cosmology: theory - galaxies: formation - galaxies:
 morphologies - galaxies: kinematics and dynamics}

\newpage

\section{Introduction}

Gravitational interactions between galaxies are believed to play a
major role in determining galaxy physical properties (Spitzer \& Baade
1951; Toomre \& Toomre 1972; Toomre 1974; Schweizer 1996; Schweizer \&
Seitzer 1992; Cole \etal 1994). For example, galactic interactions are
likely to enhance star formation rates and could explain the intense
star formation seen in IRAS galaxies. Indeed, these highly luminous
infrared galaxies often show evidence of undergoing tidal interactions
or mergers (Clements \etal 1996; Leech \etal 1994).

On theoretical grounds, models have shown that close encounters of
galaxies stimulate cloud growth and star formation. Stellar and mass
distributions in galaxies are likely to be altered by tidal
collisions, and, as a result, the morphology of a galaxy may
evolve. For example, the disks of spiral galaxies might be puffed up
by such tidal encounters, and it has been suggested (Richstone 1976)
that spiral galaxies might be converted into S0 or elliptical galaxies
by such a mechanism. Following this idea, galaxy morphologies might be
largely determined by the local properties of the environment in which
they form and evolve, since tidal collisions between galaxies occur
more frequently in denser regions of the universe.

Alternatively, galaxy morphologies may be related to intrinsic
properties of the primordial fluctuation field.  In the Cold Dark
Matter (CDM) theory in which the density parameter $\Omega_0=1$, a
bias must be invoked whereby the overdense regions contain galaxies,
and the underdense regions (voids) are deficient in luminous galaxies,
so that the luminous matter has stronger density correlations than the
underlying dark matter.  Early discussions of CDM used Gaussian
fluctuations as described by the peaks formalism (Bardeen \etal 1986)
and introduced a bias of the galaxy types with respect to mass by
identifying 3$\sigma$ primordial fluctuations with protoelliptical
galaxies and $2\sigma$ fluctuations with protospirals in order to
match the observed frequency and clustering of luminous galaxies
(Blumenthal \etal 1984; Evrard 1989; Evrard, Silk \&
Szalay 1990).  Though this approach offered the advantage of
reproducing several observed correlations, no physical mechanism was
proposed to explain this intrinsic biasing until Dekel \& Silk (1986)
argued that 1$\sigma$ fluctuations that satisfy the cooling criterion
for galaxy formation (Rees \& Ostriker 1977; Silk 1977) would be
systematically of low-mass and shallow gravitational potential wells
so that the rms density fluctuations should be especially vulnerable
to disruption by supernovae explosions.  A prediction of this model is
that an extensive distribution of 'failed galaxies', identified as
dwarf elliptical galaxies, would populate the low-density regions of
the universe.  However, it has been argued that observations strongly
constrain the hypothesis that dwarf ellipticals can account for
biasing (Bingelli 1989), the more luminous, observed dwarf ellipticals
clustering together with the bright galaxies.

An alternative physical bias mechanism may naturally involve galaxy
interactions as far as these interactions are assumed to trigger star
formation.  According to this approach, low-density regions of the
universe should contain nascent or unborn galaxies, essentially gas
clouds, since collisions are much less frequent in underdense
regions. Conversely, in denser regions, substantial interactions at
the epoch of formation would induce early star formation and
subsequent collisions may redistribute the galactic stellar content
and alter the galaxy morphology.

In this paper, we build a simple semi-analytical model of galaxy
interactions to understand galaxy morphologies and how they relate to
the properties of the environment and the fluctuation field at the
epoch of formation. We shall show that simple collision rules provide
an explanation of the morphological distribution of galaxies with
respect to environment, as well as of fundamental correlations in
their structural properties. Moreover, biasing of galaxies of
different morphological types with respect to mass originates
naturally in our scenario. The model is based on the results of N-body
simulations of tidal collisions (Aguilar \& White 1985; AG85
hereafter). The paper is organized as follows: we first describe our
model. Galaxy collisions are characterized in terms of the rate of
change of binding energy induced in a given galaxy, at a given epoch,
by tidal encounters with a set of background galaxies (\S 2).  The
cases for field and cluster galaxies are treated separately as the
physics of collisions is quite different in these two environments.
In \S 3, the total change of binding energy in a galaxy that went
through a series of tidal collisions since its formation (we shall
refer to this quantity as the 'collision factor') is computed and
characterizes the collision history of the galaxy.  Scalings of the
collision factor with local galaxy density are obtained. In \S 4, we
propose a phenomenological definition of galaxy morphological types
that can be easily expressed as conditions on the collision factor (\S
4.2). We show (\S 4.3) that these conditions are in turn conditions on
the formation redshift of galaxies, and we define redshift cuts, in
the space of formation redshifts, that delineate spiral, S0 and
elliptical galaxies. Sections 5 and 6 are devoted to normalizing our
model to various sets of observations, assuming CDM initial
conditions. Predictions of the model are given in the remaining
sections. The model predicts the redshift of formation (\S 7) as well
as the relative bias between galaxies of various morphological types
(\S 8) in the field. The morphology-density relation is well
reproduced by the model (\S 9). The predictive power of the model is
illustrated in \S 10, where the evolution of the morphological
populations with redshift is predicted, under CDM initial conditions,
in the field as well as in denser environments.  Throughout the paper,
the case for an Einstein-de Sitter ($\Omega_0=1$) with a Hubble
constant $H_0=50$ km/s/Mpc is assumed.

\section{Characterization of galaxy collisions}

We characterize galaxy collisions in terms of the rate of change of
binding energy $E$ of a galaxy colliding with an other.  This can be
expressed in a general form for a test galaxy (of mass $M$ and radius
$R$) interacting with a set of background galaxies (of mass $M_p$ and
radius $R_p$) with relative velocity $v$ as (Richstone 1975; AG85):
\begin{equation}
\label{eq1}
{\dot \Delta}=n_p v R^2 f_E(M_p/M,V/v)
\end{equation}
where ${\dot \Delta}$ stands for $d\ln E/dt$.  $n_p$ is the number
density of background galaxies, and $V$ the internal velocity
dispersion of the test galaxy.  The dimensionless function $f_E$ is
plotted in figure 8 of AG85 for equal masses galaxies. It is easy to
see from this figure that $f_E$ scales roughly as the square of the
ratio $V/v$ so that equation (\ref{eq1}) simply reads as:
\begin{equation}
\label{eq2}
{\dot \Delta}=f R^2\Big({V\over v}\Big)^2 v n_p
\end{equation}
where $f$ is the slope of the function $f_E$ and is equal to $f\approx
30$ (AG85; figure 8). For unequal galaxy masses, two modifications of
(\ref{eq1}) may be anticipated: 1) $R$ should be replaced by $R_>$,
where $R_>={\mathrm{max}}(R,R_p)$, and 2) powers of the mass ratio
$M/M_p$ may appear.  For a test galaxy $(R,M)$ perturbed by a set of
background galaxies $(R_p,M_p)$ with distribution
$dn_p=\eta(M_p,z)dM_p$, the perturbation $\delta V$ on the stellar
velocity field can be estimated from the impulse approximation
(Spitzer 1958) and straight line trajectories (provided $v\gg V$) to
be:
\begin{equation}
\label{eq3}
{{\delta V} \over V}\sim{{GM_p} \over {vV}} {R \over {p^2}}
\end{equation}
where $p$ is the impact parameter which for the dominant contribution
to the rate of energy exchange can be taken to approximately equal
$R_>$. Using the virial theorem for the test galaxy, equation
(\ref{eq3}) becomes:
\begin{equation}
\label{eq4}
{{\delta V} \over V}\sim {V \over v} { R^2 \over {R_>^2}}{M_p \over M}
\end{equation}
For equal mass galaxies, we have from (\ref{eq4}) that $\delta V/V\sim
V/v$ so that ${\dot \Delta}$ scales as $(\delta V/V)^2$ - equation
(\ref{eq2}). We assume this still holds for unequal masses with
$\delta V/V$ given by (\ref{eq4}). The rate for energy exchange now
reads as:
\begin{equation}
\label{eq5}
{\dot \Delta}=\int_{M_p} f R_>^2 \Big({V \over v}\Big)^2 \Big({M_p
\over M}\Big)^2 \Big({R \over R_>}\Big)^4 v \eta(M_p,z)dM_p
\end{equation}
where the integral is performed over the masses of background
galaxies.  As equation (\ref{eq5}) relies on the impulse approximation
and straight line trajectories, it is valid only for $v\gg V$. This
latter condition is fulfilled in galaxy clusters, for example, as the
relative velocity between galaxies is expected to be of the order of
the cluster velocity dispersion $\sigma \sim 1000$ km/s, whereas the
internal velocity dispersion of galaxies is of the order of $\sim
200-300$ km/s. In the field, the relative velocity of colliding
galaxies may be of the same order as their internal velocity
dispersion and the relative trajectories can not be approximated by
straight lines. The focusing due to mutual attraction can be easily
incorporated into equation (\ref{eq5}).  If $p$ and $v$ are the
initial impact parameter and relative velocity, energy conservation
during the collision implies that, in the frame of the reduced
particle of mass $\mu=M_pM/(M_p+M)$:
\begin{equation}
\label{eq6}
\frac{1}{2}\mu v^2=\frac{1}{2}\mu v_{col}^2-\frac{GM M_p}{R_{col}}
\end{equation}
where $v_{col}$ is the relative velocity at closest approach
$R_{col}$, and $G$ is the gravitational constant.  Conservation of
momentum per unit mass $L$ implies $L=pv=R_{col}v_{col}$ from which we
get:
\begin{equation}
\label{eq7}
p^2=R_{col}^2[1+\frac{2G(M+M_p)}{R_{col}v^2}]\approx
R_{col}^2[1+\frac{2G(M_>)}{R_{col}v^2}]
\end{equation}
Using the virial theorem for the largest of the two colliding galaxies
(of mass $M_>$ and internal velocity dispersion $V_>$), we have:
\begin{equation}
\label{eq8}
p^2\approx R_{col}^2\Big[1+\Big(\frac{V_>}{v}\Big)^2\Big]
\end{equation}
Taking $R_{col}\sim R_>$, gravitational focusing roughly amounts to
multiplying $R_>^2$ in equation (\ref{eq5}) by the factor
\begin{equation}
\label{eq9}
1+\Big(\frac{V_>}{v}\Big)^2
\end{equation}
For a similar reason, the relative velocity entering equation
(\ref{eq3}) is increased by the square root of (\ref{eq9}).  However,
the energy taken out of the relative motion to be injected into the
internal stellar motions lowers the relative motion in a similar
way. If $v\sim V$, the two effects nearly cancel each other. We do not
attempt to model these fairly complex processes, but simply use the
rate (\ref{eq5}) modified by the factor (\ref{eq9}). The scaling with
mass of the latter is needed in order to take into account the strong
attraction produced by large galaxies, and will be important for the
scaling of the rate of change of energy with mass and redshift.

We now integrate (\ref{eq5}) over the masses of the background
galaxies. We use $R_>=R$, the radius of the test galaxy, so that the
test galaxy is larger, on average, than the background galaxies.  We
define $\int M_p\eta(M_p)d M_p=\varepsilon {\bar \rho}$, where ${\bar
\rho}$ is the mean density in the universe and $\varepsilon$ is the
mass fraction in galaxies, so that we get:
\begin{equation}
\label{eq10}
{\dot \Delta}\approx f R^2 \Big({V \over v}\Big)^2 \Big({M \over {\bar
M}}\Big)^{-2} {\varepsilon {\bar \rho} v \over {\bar M}}, \ \ \ \ v\gg
V
\end{equation}
and
\begin{equation}
\label{eq11}
{\dot \Delta}\approx f R^2 \Big({V \over v}\Big)^4 \Big({M \over {\bar
 M}}\Big)^{-2} {\varepsilon {\bar \rho} v \over {\bar M}}, \ \ \ \
 v\sim V
\end{equation}
where ${\bar M}$ in equation (\ref{eq10}) and (\ref{eq11}) is defined
by:
\begin{equation}
\label{eq12}
{\bar M}\equiv{{\int M_p^2 \eta(M_p) dM_p}\over {\int M_p
\eta(M_p)dM_p}}.
\end{equation}
${\bar M}$ is dominated by the larger masses due to the $M_p$ factors:
hence, we expect ${\bar M}$ to be close to the mass of the larger
galaxies.

To summarize this section, we have obtained scaling relations for the
rate of change of binding energy as a function of the various
parameters characterizing the colliding galaxies.  Table 1 gives a
list of the names and meaning of the parameters used so far. Table 2
summarizes the results we shall build our model on in the next
sections, and the various assumptions used.\\

\section{Collision history of galaxies}

The rate of change of binding energy for a galaxy interacting with
surrounding galaxies, between redshift $z$ and $z+dz$ is
\begin{equation}
\label{eq13}
d\ln E = {\dot \Delta}{dt \over{dz}}dz,
\end{equation}
 where ${\dot \Delta}$ has been given by (\ref{eq10}) and (\ref{eq11})
in the previous section. Integrating (\ref{eq13}) over time from the
redshift of formation $z_{nl}$ of the galaxy up to redshift $z$ yields
the total increase of binding energy due to a series of collisions
experienced by the galaxy during its lifetime. This integrated
quantity characterizes what we shall call the 'collision history' of a
galaxy. To evaluate this quantity, we need to use the relevant scaling
with redshift of the various quantities entering equations
(\ref{eq10}), (\ref{eq11}) and (\ref{eq12}), which have been listed in
Table 1.  These scalings will be different in cluster and field
environments. We shall therefore distinguish from now on the cases for
cluster and field galaxies. We remind the reader that all the results
given in this paper are for an Einstein-de Sitter universe, that is
$\Omega_0=1$, and for $h=0.5$, where $h$ is the Hubble constant in
units of $100$ km/s/Mpc.

\subsection{Redshift evolution of ${\bar M}$}

By construction (equation (\ref{eq12})), ${\bar M}$ depends on the
mass distribution $\eta(M)$ of the background galaxies.  $\eta(M)$
evolves with redshift and so does the average mass ${\bar M}$ in our
model. We evaluate the scaling of ${\bar M}$ with redshift in this
section.
   
\subsubsection{Press \& Schechter mass function}

With the assumption that the initial fluctuations are Gaussian
distributed, the mass function of dark halos can be obtained from the
condition that a fluctuation is non-linear at a given mass scale, but
not at an immediately larger scale (Press \& Schechter 1974; Schaeffer
\& Silk 1985). The number density of non-linear condensations of mass
between $M$ and $M+dM$ at redshift $z$ is, in an Einstein-de Sitter
universe:
\begin{equation}
\label{eq13a}
\eta(M,z)=\sqrt{\sur{2}{\pi}}\sur{\rho_0}{M^2}
\sur{d\ln\sigma^{-1}(M)}{d\ln M}\sur{\delta_c(1+z)}{\sigma(M)}\exp
\Big[-\sur{1}{2}\Big(\sur{\delta_c(1+z)}{\sigma(M)}\Big)^2\Big]
\end{equation}
where $\rho_0$ is the present day average density of the universe and
$\delta_c$ is the linearly extrapolated threshold on the density
contrast required for structure formation. We adopt the canonical
value of the spherical model (Lema\^{\i}tre 1933; Peebles 1980; Gunn
\& Gott 1972), i.e. $\delta_c\approx 1.68$, and the average density of
an object collapsing at redshift $z$ is $178\rho_0(1+z)^3$. The
relative mass fluctuation $\delta M/M$ in a volume that contains a
mass $M$ in the linear stage enters equation (\ref{eq13a}) through its
variance
\begin{equation}
\label{eq13b}
\sigma(M)=\Big<\Big(\sur{\delta M}{M}\Big)^2\Big>^\sur{1}{2},
\end{equation}
which is known as soon as the primordial fluctuation spectrum is
specified.  In the following, we shall use the Cold Dark Matter (CDM)
spectrum that can be parameterized (e.g., Narayan \& White 1988) as a
function of the comoving scale $R$, corresponding to the mass scale
$M$, as
\begin{equation}
\label{eq13c}
\sigma_{CDM}(M)=16.3(1-0.3909R^{0.1}+0.4815R^{0.2})^{-10}/b
\end{equation}
where the bias parameter $b$ has been introduced and is specified by
the amplitude of underlying matter fluctuations at $8h^{-1}$ Mpc,
$\sigma_8$.  The parameterization (\ref{eq13c}) is given for $h=0.5$,
a value that we shall adopt throughout the paper.

At a given redshift $z$, we shall now require that gas cooling occurs
after virialization to allow for gas fragmentation and star formation
(Silk 1977, Rees \& Ostriker 1977).  This is important for evaluating
the number density of background galaxies, as only a subset of all
mass condensations counted for by the Press \& Schechter prescription
(\ref{eq13a}) are actual galaxies.

\subsubsection{The cooling constraint}

The condition for cooling to be effective is a condition on the
relative importance of the cooling time-scale and the dynamical
time-scale.  Various processes contribute to gas cooling.  Assuming
that during the collapse, the gas is shock heated and reaches virial
temperature before settling into a disk, we are led to consider only
those cooling processes that are efficient at temperatures of the
order of the virial temperature.  For simplicity, we assume that line
cooling is the dominant mechanism, and shall adopt the following
cooling function for gas temperatures of the order $\sim 10^5-10^6$ K
(e.g., Sutherland \& Dopita 1993):
\begin{equation}
\label{eq13e}
\Lambda(T)\approx 2.5\times 10^{-21} T^{-1/2} n^2\ \ \ {\mathrm erg
cm^{-3} s^{-1}}
\end{equation} 
where $n$ is the particle density of the gas within a galaxy and $T$
the virial temperature. Zero metallicity has been assumed.  The
typical cooling time-scale is then
\begin{equation}
\label{eq13f}
t_{cool}\sim 3{n kT\over \Lambda(T)}\approx 5.5\times 10^6
\Big({n\over 1{\mathrm cm^{-3}}}\Big)^{-1}\Big({T\over 10^{6} {\mathrm
K}}\Big) ^{3/2}\ \ \ {\mathrm yr}
\end{equation} 
where $k$ the Boltzmann's constant.  If the gas makes up a constant
fraction $\cal F_B$ of the total mass $M$ of the galaxy, and is
uniformly distributed within the virial radius $R$, then the gas
density $n$ is
\begin{equation}
\label{eq13h}
n=\sur{3}{4\pi \mu m_p}\sur{{\cal F_B}M}{R^3}=1.6\times 10^{-3}
\Big({M\over {10^{12} M_\odot}}\Big) \Big({R \over 100 {\mathrm
kpc}}\Big)^{-3}\Big({{\cal F_B}\over 0.1}\Big) \ \ \ {\mathrm cm^{-3}}
\end{equation}
where $m_p$ is the proton mass and $\mu$ the mean molecular weight,
that for an hydrogen-helium plasma with primordial abundances is
$\mu\approx 0.6$.  The temperature of the gas is obtained from the
viral equation:
\begin{equation}
\label{eq13i}
kT\approx \mu m_p {V^2\over 3} \approx {\mu m_p}{GM\over {5R}}
\end{equation}
where $V$ is the 3-D galaxy velocity dispersion.  In (\ref{eq13i}), we
have assumed that the galaxy is spherical and the gas is homogeneously
distributed within $R$.  Plugging (\ref{eq13h}) and (\ref{eq13i}) into
(\ref{eq13f}), we obtain the following cooling time-scale:
\begin{equation}
\label{eq13j}
t_{cool}\approx 1.7\times 10^9 \Big({{\cal F_B}\over 0.1}\Big)^{-1}
\Big({M\over {10^{12}\ M_\odot}}\Big)
^{1/2}\Big({R\over{100{\mathrm kpc}}}\Big)^{3/2}\ \ \ {\mathrm yr}
\end{equation}
The dynamical time-scale can be estimated from the time taken for the
galaxy to collapse after turn around:
\begin{equation}
\label{eq13k}
t_{dyn}\approx{\pi\over 2}\sqrt{{R^3_{ta}\over {2GM}}}\approx
1.5\times 10^9 \Big({M\over {10^{12}\ {\mathrm M_\odot}}}\Big)
^{-1/2}\Big({R\over{100{\mathrm kpc}}}\Big)^{3/2}\ \ \ {\mathrm yr}
\end{equation}
where $R_{ta}\approx 2 R$ is the turn around radius, and $G$ the
gravitational constant.  For efficient star formation, we require
(\ref{eq13j}) to be smaller than (\ref{eq13k}) implying:
\begin{equation}
\label{eq13l}
M<M_*\approx 9\times 10^{11}\ \ \ {\mathrm M_\odot}
\end{equation}
where $M_*$ is a critical mass whose value is fixed by the physics of
cooling.

\noindent
For the purpose of evaluating the number density of background
galaxies, we shall use the constraint (\ref{eq13l}) that gives an
upper limit of the mass of galaxies.  In figure 1, we plot the average
mass ${\bar M}$, computed from (\ref{eq13a}), as a function of
redshift (solid line) when the cooling constraint is applied.
Integrals entering (\ref{eq12}) have been computed using a minimum
mass $M_{inf}=10^{10}\ M_\odot$.  For comparison, ${\bar M}$ obtained
from (\ref{eq13a}) with no cooling constraint is also plotted (dashed
line).  When the cooling constraint is used, the average mass ${\bar
M}$ exhibits only a weak dependence on redshift over the range $z\lta
3-4$, much weaker than when no cooling is used.  Moreover, the average
mass is found to be of the order of $M_*$. We shall adopt this value
hereafter and take ${\bar M}$ constant with redshift for $z\lta 3-4$.

\subsection{Field galaxies}

The average density of the universe in equation (\ref{eq11}) scales
with redshift as
\begin{equation}
\label{eq14}
{\bar \rho}=\rho_0(1+z)^3
\end{equation}
where $\rho_0$ is the average density of the universe at the present
epoch.  We assume that the average density in galaxies reflects the
density of the universe at the epoch of formation $z_{nl}$ within a
universal factor (that depends on the properties of the collapse). We
then can relate the radius $R$ of the test galaxy to its mass $M$ as
\begin{equation}
\label{eq15}
R=R_*\Big({M\over M_*}\Big)^{1/3}\Big({{1+z_{nl}}\over
{1+z_*}}\Big)^{-1}
\end{equation}
For simplicity, we have expressed the relevant quantities for the test
galaxy in terms of the same quantities for the typical $M_*$ galaxy
introduced in the previous section.  A typical $M_*$ galaxy has radius
$R_*$, mass $M_*$, internal velocity dispersion $V_*$, and form at
redshift $z_*$.  Using equation (\ref{eq15}) and the virial theorem,
we obtain the following scaling of the internal velocity dispersion of
the test galaxy $V$ with redshift:
\begin{equation}
\label{eq16}
V=V_*\Big({M \over M_*}\Big)^{1/3}
\Big({{1+z_{nl}}\over{1+z_*}}\Big)^{1/2}
\end{equation}
We finally need to know how the relative velocity $v$ evolves with
redshift. We assume that the mean relative velocity of colliding
galaxies can be inferred from the peculiar motions of galaxies in the
linear regime at redshift $z$:
\begin{equation}
\label{eq17}
v=a{{dx} \over {dt}}=v_0(1+z)^{-1/2}
\end{equation}
where $x$ is a comoving coordinate, $a$ is the scale factor of the
universe, and $v_0$ the present-day galaxy peculiar velocity in the
field.  This assumption is valid at high $z$ in the linear regime.  An
accurate scaling at low redshift should take into account pairwise
velocity on small scales in the non-linear regime; however,
(\ref{eq17}) introduces only a weak dependence on redshift, and we
shall adopt this scaling in the following. Plugging equations
(\ref{eq14}),(\ref{eq15}),(\ref{eq16}) and (\ref{eq17}) into equation
(\ref{eq11}), we get
\begin{equation}
\label{eq18}
{\dot \Delta}= \varepsilon f R_*^2{v_0 \rho_0 \over M_*} \Big({ V_*
\over v_0}\Big)^4 (1+z)^{9/2}.
\end{equation}
The redshift integration of equation (\ref{eq18}) from the redshift of
formation $z_{nl}$ of the test galaxy then yields the total increase
of binding energy due to collisions, up to an epoch characterized by
redshift $z$.  In the following, we shall call this quantity the
'collision factor' and denote it $\Delta$. In an Einstein-de Sitter
universe, redshift and time are related through:
\begin{equation}
\label{eq19}
dt=-H_0^{-1}(1+z)^{-5/2}dz
\end{equation}
where $H_0$ is the present day value of the Hubble constant.
Integration of (\ref{eq18}) over redshift thus leads to:
\begin{equation}
\label{eq20}
\Delta^{field}(z)\equiv \int_{z_{nl}}^{z}{\dot
\Delta}\sur{dt}{dz}dz=\Delta_*(1+z_{nl})^3 \Big[1- \Big({{1+z}\over
{1+z_{nl}}}\Big)^3\Big]
\end{equation}
where $\Delta_*$ is the dimensionless constant
\begin{equation}
\label{eq21}
\Delta_*=\varepsilon(f/3)R_*^2\Big({ V_* \over v_0}\Big)^4{v_0 \over
H_0} {\rho_0 \over M_*}.
\end{equation}
A numerical estimate of (\ref{eq21}) can be obtained by fixing the
parameters describing a typical $M_*$ galaxy today, and the relative
velocity $v_0$.

\subsection{Cluster Galaxies}

For field galaxies, collisions are only important at early times when
the density is large. Similarly, we expect little effect in the lower
density regions of clusters. In the very high density regions of rich
clusters, however, the energy exchange due to collisions is high and
is given by equation (\ref{eq10}) with ${\bar \rho}$ replaced by the
local density $\rho$ and $v$ by the local galaxy velocity dispersion
of the cluster.  Also, in the focusing factor (\ref{eq9}), the term
$(V_>/v)^2)$ that was dominant in the field can now be neglected.
Using equations (\ref{eq15}) and (\ref{eq16}), equation (\ref{eq10})
thus becomes:
\begin{equation}
\label{eq22}
\dot \Delta=f \sur{\epsilon \rho v}{M_*} R_*^2
\Big(\sur{V_*}{v}\Big)^2 \Big(\sur{M}{M_*}\Big)^{-2/3}
\Big(\sur{1+z_{nl}}{1+z_*}\Big)^{-1}.
\end{equation}
Performing the same integration over redshift as in the previous
subsection, now for cluster environment, yields the change of biding
energy for a cluster galaxy from the epoch $z_{nl}$ up to redshift
$z$:
\begin{equation}
\label{eq23}
\Delta^{cluster}(z)=\Delta_c (1+z_*)^{-3/2} \Big({M\over
M_*}\Big)^{-2/3}\Big({V_*\over v}\Big) \Big({\rho \over \rho_0}\Big)
\Big({{1+z_{nl}}\over {1+z_*}}\Big)^{-5/2} \Big[\Big({{1+z}\over
{1+z_{nl}}}\Big)^{-3/2}-1\Big]
\end{equation}
with
\begin{equation}
\label{eq24}
\Delta_c={2\over 3}{f\over H_0} R_*^2 {\varepsilon \rho_0 \over M_*}
V_*,
\end{equation}
where the average density of the universe, $\rho_0$, has been
introduced for convenience.
\subsection{Scaling of $\Delta^{cluster}$ with density}
In equation (\ref{eq23}), both $\rho$ and $v$ can be related to the
local galaxy number density $n$, provided a model is assumed for the
density distribution of the cluster. Let us consider for instance the
following cluster profile:
\begin{equation}
\label{eq25}
\rho(R)=\rho_c\Big({R \over R_c}\Big)^{-p}
\end{equation}
where $\rho_c$ is the cluster density at a radius $R_c$ from the
center, and depends on the mass and redshift of formation of the
cluster. By use of the virial theorem and (\ref{eq25}), the velocity
dispersion $v$ can be expressed as a function of the cluster density
$\rho$:
\begin{equation}
\label{eq26}
v=v_c\Big({\rho_c \over \rho_0}\Big)^{1\over p} \Big({\rho\over
\rho_0}\Big)^{{p-2}\over{2p}}
\end{equation}
where $v_c$ is the cluster velocity dispersion at radius $R_c$. As
expected for $p=2$, the velocity dispersion of a given cluster is
constant and independent of the density $\rho$. This velocity,
however, is not the same for all clusters since it depends on
$\rho_c$:
\begin{equation}
\label{eq27}
v=v_c\Big({\rho_c \over \rho_0}\Big)^{1/2}\ \ \ p=2.
\end{equation}
The collision factor (\ref{eq23}) scales, in this simple case, as
\begin{equation}
\label{eq28}
\Delta^{cluster}\propto \rho \rho_c^{-1/2}.
\end{equation}
We can relate the cluster density $\rho$ to the local galaxy number
density $n$ by assuming that galaxies trace mass, that is $n \propto
\rho$. We shall however adopt a somewhat more realistic model for the
cluster density profile, namely an isothermal $\beta$ model (Cavaliere
\& Fusco-Femiano 1976) in which the galaxy number density in the
cluster is related to the gas density profile $\rho_g(R)$ by:
\begin{equation}
\label{eq29}
\rho_g(R)\propto n^\beta (R)
\end{equation}
where $\beta$ is a fitting parameter to the projected X-ray surface
brightness of the cluster (Jones \& Forman 1984). Assuming that the
gas makes up a fraction of the total mass that is roughly constant
throughout the cluster,
we have for the cluster density $\rho$:
\begin{equation}
\label{eq30}
\rho(R)\propto \rho_g(R) \propto n^\beta (R)\propto \Big({R\over
R_c}\Big)^{ -3\beta}\ \ \ \ R\gg R_c.
\end{equation}
At large radius, $\rho$ behaves with radius as equation (\ref{eq25})
with $p=3\beta$.  Expressed in terms of the galaxy number density,
equation (\ref{eq28}) now scales, for collisions occurring at $R>R_c$,
as:
\begin{equation}
\label{eq31}
\Delta^{cluster}\propto n^\beta n_c^{-\beta/2}
\end{equation}
where $n_c$ is the galaxy number density at radius $R_c$.  We are led
to two different cases: 1) collisions at a fixed density $n$ within a
given cluster, and 2) collisions at a fixed density averaged over a
set of clusters. The scaling with density of the collision factor
(\ref{eq31}) will be quite different in these two cases. It is of
interest of considering case 2) for the purpose of comparing the
predictions of our model to observations, which we will present later
in this paper.  For example, (\ref{eq31}) can be averaged, at a fixed
galaxy number density $n$, over a set of clusters with varying 'central'
density $n_c$. Bahcall (1979) shows that the distribution of Abell
clusters decreases as a function of luminosity $L$ according to a
Schechter law
\begin{equation}
\label{eq32}
\Phi(L)\propto L^\alpha\exp \Big(-L/L_a\Big)
\end{equation}
where $L_a$ is the cluster optical luminosity within the Abell radius
$R_a$. Assuming that the cluster optical luminosity is proportional to
the galaxy number density, the central density distribution of Abell
clusters is
\begin{equation}
\label{eq33}
\Phi(n_c)\propto n_c^\alpha \exp\Big(-n_c/n_a\Big)
\end{equation}
with $n_a\propto L_a/R_a^3$. The collision factor (\ref{eq31})
averaged at fixed $n$ over the distribution (\ref{eq33}) for $n_c\ge
n$ is then
\begin{equation}
\label{eq34}
<\Delta^{cluster}>\sim n^\beta <n_c^{-\beta/2}>
\end{equation}
where $<>$ denotes average over (\ref{eq33}):
\begin{equation}
\label{eq35}
<n_c^{-\beta/2}>=n_a^{-\beta/2}f(n/n_a)
\end{equation}
where the function $f(x)$ is defined by:
\begin{equation}
\label{eq36}
f(x)=\Big[{\int_x^\infty t^{\alpha-\beta/2} \exp(-t)dt
\over{\int_x^\infty t^\alpha \exp(-t)dt}}\Big].
\end{equation}
$f(n/n_a)$ is dominated by the contribution of densities $n\sim n_a$
and behaves roughly as $(n/n_a)^{-\beta/2}$. Thus, (\ref{eq34})
becomes
\begin{equation}
\label{eq37}
<\Delta^{cluster}>\sim n^{\beta/2}.
\end{equation}
In the following, we will use equation (\ref{eq37}) for the purpose of
evaluating the morphology-density relation in our model.
\subsection{Comparison between the field and clusters}
Several differences in the collision factor in the two environments
considered originate from the different scalings used in these two
cases:
\begin{itemize}
\item In the field, the collision factor depends essentially on the
present day density of the universe, whereas in clusters, $\Delta$ is
a function of the local density $\rho$. As a consequence, the energy
exchange in clusters are generally much more important than in the
field, and depend on the environment, since they increase with
increasing density.
\item In the field, collisions at high redshift are more efficient (in
terms of the fractional rate of change of binding energy per collision
(\ref{eq18})) than recent collisions. This is a direct consequence of
the strong increase of the average density of the universe with
redshift. This effect overseeds the fact that collisions for galaxies
with small radii (that is, at a given mass, those formed at high
$z_{nl}$ - equation (\ref{eq15})), are less efficient than for large
galaxies.  Thus, at a given mass, (\ref{eq20}) implies that the higher
the redshift of formation of a galaxy, the higher the collision
factor.  On the contrary, in a given cluster, the fractional rate of
change of binding energy per collision (\ref{eq22}) is constant in
time, as soon as the cluster has virialized.  At a given mass, the
higher the redshift of formation of the galaxy, the lower the galaxy
radius (\ref{eq15}), and the lower the collision factor (\ref{eq23}).
\end{itemize}
 
\section{A phenomenological definition of Hubble types}

\subsection{Angular momentum and morphological types}

A key phenomenon to understand when elaborating a consistent picture
of galaxy formation is the angular momentum history of galaxies of
various Hubble types. It is believed that the angular momentum of
primeval galaxies originates from tidal torques induced by the
presence of neighboring galaxies (Doroshkevich 1970; White 1984).  At
the early epochs of galaxy formation, protogalaxies indeed interact
strongly with their surroundings. The angular momentum acquired at
that time may have a definitive impact on galaxy morphologies.  It
would be however misleading to associate isolated structures, i.e.,
objects that are less subject to tidal torquing, with elliptical
galaxies. Following this argument, one would then expect to find fewer
low-spin galaxies in high density environments, a prediction in
disagreement with the observation of an increasing fraction of
early-types when one goes towards richer clusters (Dressler 1980).

It is unlikely that the large amount of spin needed to rotationally
support fully grown disk galaxies may be explained solely by the
primordial angular momentum (e.g., Barnes \& Efstathiou 1987).
Dissipative collapse of the radiatively cooled gas in a dark halo must
be invoked.  Assuming that the gas angular momentum is conserved
during the collapse phase, one needs a collapse by a factor 10 to
achieve rotational support (Fall \& Efstathiou 1980). A large initial
angular momentum is not a sufficient condition for the formation of
disk galaxies, as various processes, such as tidal interactions
occurring during the lifetime of a galaxy, may induce angular momentum
transport from the galaxy center to the outer parts. Thus an initially
rapidly rotating protogalaxy could end up as an elliptical galaxy.
Recent simulations indeed find an extremely high efficiency of angular
momentum transfer resulting in disks that are far too small.  Even if
the gas starts to sink towards the galactic center, the settlement
into a disk structure requires a gentle infall that could be perturbed
by the tidal field. Indeed, modeling of disk formation as well as
chemical evolution models (Lacey \& Fall 1983; Rocca-Volmerange \&
Schaeffer 1990) supports slow disk formation via gas infall from a
pre-existing halo.  Spirals are found to have a relatively constant
rate of star formation over the past 10 Gyr, and infall provides a
possible means of regulating the gas supply and maintaining the disk
in a state of marginal gravitational instability (Sellwood \& Carlberg
1984). However, as soon as a protogalactic cloud collapses and
decouples from the universal expansion, collisions are expected to
occur due to its relative velocity with respect to other
clouds. Virialization should be effective within one or a few Hubble
times after formation and collisions during this period are not
expected to greatly modify the final structure. More recent
collisions, however, would inhibit the gentle infall of the gas needed
to allow for the formation of a disk.  Clearly, the collision history
over the whole life of a galaxy since its epoch of formation should be
taken into account if one wants to predict its morphological type.

\subsection{Definition of Hubble types}

The previous discussion leads us to define, in the framework of our
model, the morphological types as follows: the condition for a spiral
galaxy to form out of a cloud that first became non-linear at a
redshift $z_{nl}$ is to experience few, if any, collisions between the
epoch of its formation and the epoch under consideration.  Strong
collisions will, on the contrary, prevent the gas from settling into a
disk and allow for tidal exchanges that average out angular
momentum. We shall consequently assume that a sizeable number of
collisions between $z_{nl}$ and the epoch characterized by $z$ leads
to the formation of elliptical galaxies.  These rules for the
formation of morphological types can be expressed in terms of
conditions on the collision factor $\Delta$, for field and/or cluster
galaxies.  At any redshift $z$, the collision factor for a spiral
galaxy is assumed to satisfy the following condition:
\begin{equation}
\label{eq38}
\Delta(z)<\Delta_{th}^{spi}\ \ \ ({\mathrm Spiral})
\end{equation} 
where $\Delta_{th}^{spi}$ is a redshift independent threshold on the
collision factor and is a free parameter in this model. An elliptical
galaxy will be such that:
\begin{equation}
\label{eq39}
\Delta(z)>\Delta_{th}^{ell}\ \ \ ({\mathrm Elliptical})
\end{equation}
where $\Delta_{th}^{ell}$ is another free parameter of our theory such
that $\Delta_{th}^{ell}>\Delta_{th}^{spi}$. Finally, galaxies whose
collision factors satisfy
\begin{equation}
\label{eq40}
\Delta_{th}^{spi}<\Delta(z)<\Delta_{th}^{ell}\ \ \ ({\mathrm S0})
\end{equation}
will define S0 galaxies in our model.  Condensed objects, either
having undergone few collisions, or which have collapsed too recently
to form a disk, may be identified with damped Ly$\alpha$ clouds,
Ly$\alpha$ forest or metal-line absorbers.

\subsection{Conditions on the formation redshift of a galaxy}

From the modeling of sections \S 3.2 and 3.3 (equations (\ref{eq20})
and (\ref{eq23})), we now rewrite conditions (\ref{eq38})-(\ref{eq40})
as conditions on the formation redshift of galaxies.

\subsubsection{Field galaxies}

In the field limit, conditions (\ref{eq38})-(\ref{eq40}) become:
\begin{equation}
\label{eq41}
1+z_{nl}<1+z^f_{spi}(z)\ \ \ ({\mathrm Spiral}),
\end{equation}
\begin{equation}
\label{eq42}
1+z_{nl}>1+z^f_{ell}(z)\ \ \ ({\mathrm Elliptical})
\end{equation}
and
\begin{equation}
\label{eq43}
1+z^f_{spi}(z)<1+z_{nl}<1+z^f_{ell}(z)\ \ \ ({\mathrm S0}),
\end{equation}
where $z^f_{spi}$ and $z^f_{ell}$ are limiting redshifts obtained by
setting $\Delta$ equal to $\Delta_{th}^{spi}$ and $\Delta_{th}^{ell}$,
respectively, in equation (\ref{eq20}). At $z=0$ (the present epoch),
and provided $z_{nl}\gg 0$, we find from (\ref{eq20}) that
\begin{equation}
\label{eq44}
1+z^f_{spi}(z=0)\approx \Big({\Delta_{th}^{spi}\over
\Delta_*}\Big)^{1/3}
\end{equation}
and
\begin{equation}
1+z^f_{ell}(z=0)\approx \Big({\Delta_{th}^{ell}\over
\Delta_*}\Big)^{1/3}.
\label{eq45}
\end{equation}
In general, $z^f_{spi}$ and $z^f_{ell}$ are functions of the redshift
$z$.

\subsubsection{Cluster galaxies}

Similar conditions on the redshift of formation can be obtained in the
cluster limit from (\ref{eq23}) and (\ref{eq38})-(\ref{eq40}),
provided that these conditions also hold in clusters.  Those galaxies
with very large radii have stronger collisions and become
ellipticals. Their redshift of formation, which at a given mass
specifies their radius (\ref{eq15}), is constrained by:
\begin{equation}
\label{eq46}
z_{nl}<z^c_{ell}(M,\rho)\ \ \ ({\mathrm Cluster\ elliptical}).
\end{equation}
Spiral galaxies must be formed at redshift
\begin{equation}
\label{eq47}
z^c_{spi}(M,\rho)<z_{nl}\ \ \ ({\mathrm Cluster\ spiral})
\end{equation}
and S0 galaxies have a redshift of formation satisfying
\begin{equation}
\label{eq48}
z^c_{ell}(M,\rho)<z_{nl}<z^c_{spi}(M,\rho)\ \ \ ({\mathrm Cluster\
S0}).
\end{equation}
It is to be noticed that, contrary to the conditions in the field,
cluster ellipticals form at a somewhat lower redshift than
spirals. This is due to the competition between two effects: galaxies
that formed earlier will have experienced more collisions by today,
but recently formed galaxies will have larger radii and the efficiency
of collisions will be higher. In clusters, the latter effect
predominates as discussed in section \S 3.5: at a given mass, recently
formed cluster galaxies have a higher collision factor than older
ones.  The limiting redshifts $z^c_{spi}$ and $z^c_{ell}$ depend on
the redshift $z$ through (\ref{eq23}).  At the present time ($z=0$),
and provided $z_{nl}\gg 0$, we have from equation (\ref{eq23}), with
$\Delta^{cluster}$ equal to $\Delta_{th}^{spi}$ and
$\Delta_{th}^{ell}$ respectively:
\begin{equation}
\label{eq49}
1+z^c_{spi}(z=0)\approx (1+z_*)\Big({M \over M_*}\Big)^{-2/3}
\Big[{\Delta_c \over \Delta_{th}^{spi}} {\rho \over \rho_0} \Big({v
\over V_*}\Big)^{-1}\Big]^{-1}
\end{equation}
\begin{equation}
\label{eq50}
1+z^c_{ell}(z=0)\approx (1+z_*)\Big({M \over M_*}\Big)^{-2/3}
\Big[{\Delta_c \over \Delta_{th}^{ell}} {\rho \over \rho_0} \Big({v
\over V_*}\Big)^{-1}\Big]^{-1}
\end{equation}
Note that $\Delta_{th}^{ell}$ and $\Delta_{th}^{spi}$ in (\ref{eq49})
and (\ref{eq50}) are not necessarily the same thresholds as in the
field.

\subsection{Summary}

We have presented a model for galaxy collisions. Using a
phenomenological definition of galaxy morphological types, we have
obtained, for each type in the limit $z=0$, a condition on the
redshift of formation of a galaxy: equations (\ref{eq41})-(\ref{eq43})
for field galaxies, (\ref{eq46})-(\ref{eq48}) for cluster galaxies.
These conditions depend on the thresholds $\Delta_{th}^{ell}$ and
$\Delta_{th}^{spi}$, which, as well as $z_*$, are free parameters of
the model and scale all of the model predictions.

\section{Determination of $z_*$}

The conditions derived above on formation redshifts are all
independent of the initial spectrum of fluctuations, which need not be
specified
\footnote{We have introduced the CDM spectrum in \S 3.1 for the sole
purpose of studying the variation of the mean galaxy mass ${\bar M}$
with redshift.}. Once an initial fluctuation spectrum is chosen,
fluctuations at a given mass scale $M$ are characterized by their
effective height $\nu(M)$ in the matter density field linearly
extrapolated until the present epoch:
\begin{equation}
\label{eq51}
\nu(M)\equiv {\delta_c (1+z_{nl}) \over \sigma(M)}
\end{equation}
where $\delta_c$ is the threshold on the linear density contrast
required in structure formation theories, $\sigma$ the linear variance
of mass fluctuations on scale $M$ - equation (\ref{eq13b}), and
$z_{nl}$ the redshift at which fluctuations on scale $M$ virialize.
The mass function (\ref{eq13a}) can then be expressed in terms of the
height $\nu$, given by (\ref{eq51}), as:
\begin{equation}
\label{eq54}
\eta(M)=-\sqrt{\sur{2}{\pi}}\sur{\rho_0}{M^2} \sur{d\ln\sigma(M)}{d\ln
M}\nu(M)\exp \Big[-\sur{1}{2}\nu^2(M)\Big].
\end{equation}
 
There is a consistency relation in order for the cooling constraint
(\ref{eq13l}) that prevails at $M\approx M_*$ to provide the correct
galaxy luminosity function $\Phi(L)$.  The local luminosity function
of the Stromlo-APM redshift survey (Loveday \etal 1992) is well fitted
by a Schechter function:
\begin{equation}
\label{eq53}
\Phi(L)=\Phi_*\Big({L\over L_*})^{-\alpha}\exp(-L/L_*)\sur{dL}{L_*},
\end{equation}
where $\alpha\approx 0.97$ and $\Phi_*\approx 1.75\times 10^{-3}$
Mpc$^{-3}$ (for $h=0.5$).  Assuming no luminosity evolution, we
require that (\ref{eq53}) and (\ref{eq54}) match on the scale $M=M_*$,
that is $\Phi(L)dL|_*=\eta(M)dM|_*$. We get:
\begin{equation}
\label{eq55}
\nu_*\exp(-\sur{\nu_*^2}{2})\approx -\sur{\Phi_*}{\exp(1)}\sqrt{\sur{\pi}{2}}
\sur{M_*}{\rho_0}\sur{d\ln M}{d\ln\sigma(M)}\Big|_{M=M_*}
\end{equation}
where $\nu_*\equiv \delta_c(1+z_*)/\sigma(M_*)$ and we have assumed a
constant mass-to-light ratio. Solving equation (\ref{eq55}), we find
$\nu_*\approx 2.8$. Using CDM initial conditions (\ref{eq13c}) with a
bias parameter $b=1.67$ ($\sigma_8\approx 0.6$) required to reproduce
the abundance of the observed X-ray clusters at present in an
$\Omega_0=1$ universe (White, Efstathiou \& Frenk 1993), we have
$\sigma(M_*)\approx 3.3$, from which we infer
\begin{equation}
\label{eq56}
(1+z_*)\approx 5.5.
\end{equation}
Once we know $z_*$, the numerical values of the parameters of our
model can be fixed. If we assume spherical symmetry, the radius of an
$M_*=10^{12}\ M_\odot$ galaxy formed at redshift $z_*$ is
$R_*=(3M_*/4\pi \rho_0 178 (1+z_*)^3)^{1/3}\approx 50$ kpc. Its
velocity dispersion is $V_*^2\approx 0.5 GM_*/R_*$, implying
$V_*\approx 200$ km/s.  We can then compute the values of $\Delta_*$
and $\Delta_c$ (equations (\ref{eq21}) and (\ref{eq24})) provided we
fix the quantities $v_0$ and $\varepsilon \rho_0/M_*$.  The mean
number density of $M_*$ galaxies today, $\varepsilon\rho_0/M_*$, is
inferred from the local luminosity function of the Stromlo-APM survey,
given by Loveday \etal (1992): $\varepsilon\rho_0/M_*\approx \Phi_*
\approx 1.75\times 10^{-3} $ Mpc$^{-3}$. Taking $v_0=200$ km/s, we
find $\Delta_*=1.75\times 10^{-4}$ and $\Delta_c=3.5\times
10^{-4}$. The numerical values of the parameters of the model are
summarized in table 3.  The only remaining quantities to be determined
in order to fully normalize our model are $\Delta_{th}^{spi}$ and
$\Delta_{th}^{ell}$.

\section{Determination of $\Delta_{th}^{spi}$ and $\Delta_{th}^{ell}$}

The quantities $\Delta_{th}^{spi}$ and $\Delta_{th}^{ell}$ are fixed
by requiring that the model produces the observed fractions of
morphological types in the field today. These fractions can be
evaluated in the following manner. Our phenomenological definition of
morphological types implies definite conditions on the redshift of
formation of galaxies. For example, field ellipticals are required to
form at a redshift $z_{nl}\ge z^f_{ell}$ (equation (\ref{eq42})) in
order to have experienced substantial energy exchange during
collisions.  The number density of objects of mass $M$ which were
already non-linear, with a mass between $M/\lambda$ and $M$, at a
redshift $z'$ such that $z_{nl}>z'>z$ is given (Lacey \& Cole 1993)
by:
\begin{equation}
\label{eq57}
\eta(M,z,z')=\sqrt{\sur{2}{\pi}}\sur{\rho_0}{M^2}
\sur{d\ln\sigma^{-1}(M)}{d\ln M}\sur{\delta_c(1+z)}{\sigma(M)}\exp
\Big[-\sur{1}{2}\Big(\sur{\delta_c(1+z)}{\sigma(M)}\Big)^2\Big]\times
{\rm erfc}(x)
\end{equation}
where
\begin{equation}
\label{eq58}
x=\sur{\delta_c(z'-z)}{\sqrt{2}\sqrt{\sigma^2(M/\lambda)-\sigma^2(M)}}
\end{equation}
and
\begin{equation}
\label{eq59}
{\mathrm erfc}(x)={2 \over \sqrt{\pi}}\int_x^\infty \exp({-u^2})du.
\end{equation} 
The erfc term represents the probability that an object present at
$z'$ still exists at redshift $z$. The present day fraction of
ellipticals is thus given for $z'=z^f_{ell}$ by:
\begin{equation}
\label{eq60}
{\cal F}_{ell}=\sur{\int_{M_{inf}}^{M_{sup}} \eta(M,0,z^f_{ell})dM}
{\int_{M_{inf}}^{M_{sup}} \eta(M,0,0)dM}\ \ \ ({\mathrm Field\
ellipticals\ today}).
\end{equation}
The denominator of equation (\ref{eq60}) is the number density of all
condensed objects with mass between $M_{inf}$ and $M_{sup}$, while the
numerator counts only those objects that were formed before or at
$z^{ell}$ and that survived until the present epoch.  Similarly, the
fractions of spiral and S0 galaxies are given by
\begin{equation}
\label{eq61}
{\cal F}_{spi}=\sur{\int_{M_{inf}}^{M_{sup}}
[\eta(M,0,0)-\eta(M,0,z^f_{spi})]dM} {\int_{M_{inf}}^{M_{sup}}
\eta(M,0,0)dM}\ \ \ ({\mathrm Field\ spirals\ today}),
\end{equation} 
and
\begin{equation}
\label{eq62}
{\cal F}_{spi}=\sur{\int_{M_{inf}}^{M_{sup}} [\eta(M,0,z^f_{spi})-
\eta(M,0,z^f_{ell})]dM} {\int_{M_{inf}}^{M_{sup}} \eta(M,0,0)dM}\ \ \
({\mathrm Field\ S0s\ today}).
\end{equation}
Equations (\ref{eq60})-(\ref{eq62}) depend on $\Delta_{th}^{spi}$ and
$\Delta_{th}^{ell}$ through the redshift cuts $z^f_{spi}$ and
$z^f_{ell}$. The observed fractions of morphological populations in
the field today are $\approx 65$\% spirals, $\approx 10$\%
ellipticals, and $\approx 25$ \% S0s (Dressler 1980; Postman \& Geller
1984).  For the purpose of evaluating (\ref{eq60})-(\ref{eq62}), we
use $\lambda=2$ in equation (\ref{eq58}) and $M_{inf}=10^{10} \
M_\odot$. $M_{sup}$ is given by the cooling constraint
(\ref{eq13l}). The CDM spectrum (\ref{eq13c}) is used.  We find in
order to produce the correct abundance of morphological types that
$\Delta_{th}^{spi}\approx 0.003$ and $\Delta_{th}^{ell}\approx 0.01$.
The collision factor for field ellipticals is thus typically larger by
a factor at least 3 than the one for ellipticals.

\section{Redshift of formation of the morphological types in the field}

Equations (\ref{eq41})-(\ref{eq43}) give the typical redshift of
formation of galaxies of different morphological types.  Using the
values derived previously for the thresholds on the collision factor,
we find that typical field ellipticals form at $z_{nl}\gta 2.5$ while
typical field spirals form at somewhat smaller redshifts $z_{nl}\lta
1.5$.  S0s form, in our scenario, at intermediate redshifts.  Figure 2
illustrates this prediction. The number density of the three galaxy
types in the field today ($z=0$) are plotted as a function of the
redshift of formation of the galaxies.  The results have been obtained
from equations (\ref{eq60})-(\ref{eq62}) where the division by the
total integrated galaxy number density has been omitted.  The redshift
cuts (\ref{eq41})-(\ref{eq43}) have been used. The plot illustrates
the main features of our phenomenological model: today's field
ellipticals were formed at high redshift so as to experience efficient
energy exchange through collisions.  No galaxy forming at redshift
lower than $z\simeq 2.5$ will end up as an elliptical by today.
Conversely, recently formed galaxies have suffered little tidal
disturbance and constitute today's spirals. Finally, galaxies that
were born at intermediate redshifts define the present day S0s.

\section{Biasing between galaxy populations}

Each of the conditions (\ref{eq41})-(\ref{eq43}) implies a precise value
of the height $\nu(M)$ of fluctuations condensing out of the
primordial density field into galaxies.  For example, we infer from
(\ref{eq42}) and (\ref{eq51}) that field ellipticals of mass $M$ will
condense out of linear fluctuations of height $\nu$ such that
\begin{equation}
\label{eq63}
\nu>\nu_{ell}(M)\equiv {1.68(1+z^f_{ell})\over \sigma(M)}
\end{equation}
where we define the threshold $\nu_{ell}$ for ellipticals.  For field
spirals of mass $M$, we have
\begin{equation}
\label{eq64}
\nu<\nu_{spi}(M)\equiv {1.68(1+z^f_{spi})\over \sigma(M)}.
\end{equation}
where we define the threshold $\nu_{spi}$ for spirals.  For $M_*$
galaxies, we have $\nu_{ell}\approx 3$ and $\nu_{spi}\approx 2$.
Equations (\ref{eq63}) and (\ref{eq64}) thus introduce a bias by
effectively requiring that field ellipticals, spirals and S0s
correspond to definite subsets of all mass condensations.  By contrast
with early theories of biased galaxy formation where $\nu$ was free to
be chosen, the formation threshold in our model results from the
modeling of the physical processes involved during galaxy formation.

\section{The predicted morphology-density relation}

As early as 1958, Abell noted that elliptical galaxies are more
frequently found in the cores of dense clusters (Abell 1958; Morgan
1961) while spiral galaxies predominate in the outer parts of clusters
and make up almost $70\%$ of field galaxies. Dressler (1980) first
noticed the existence of a relation between galaxy morphological types
and the (projected) local density in which galaxies are found. Postman
\& Geller (1984) and Giovanelli, Haynes \& Chincarini (1986) have
extended Dressler's work to assess the existence of a morphology-3-D
density relation in less dense environments such as for field galaxies
and groups.  Whether such a relation can be attributed to local
properties of the environment, as claimed by Dressler, or can be
related to initial conditions at birth (the 'nature versus nurture'
controversy) has been an important issue assessed in more recent works
(Whitmore \& Gilmore 1991; Whitmore, Gilmore \& Jones 1993).  Several
effects have been proposed that could induce evolution of the
morphology of galaxies.  Spitzer \& Baade (1951) pioneered such an
approach by suggesting that spirals could evolve into S0 galaxies by
removal of their gas content.  The dense cores of rich clusters
provide an ideal environment: in these regions, ram-pressure stripping
of the spiral interstellar medium by the intracluster gas is likely to
occur (Gunn \& Gott 1972). Direct mergers of disks (Mamon 1992) or
tidal collisions (Spitzer \& Baade 1951) are other likely
effects. High resolution simulations suggest that compact sub-$L_*$
S0s may form by the cumulative effect of tidal interactions on induced
star formation and mass loss, so-called galaxy harassment (Moore \etal
1996).  Competing theories have been developed by the advocates of the
'nature' hypothesis (Evrard \etal 1990).  These various
approaches suffer however from a number of problems: while the Evrard
\etal (1990) model fails to reproduce the observed morphology-density
relation for S0 and spiral galaxies, the result of removing the gas
content of spiral galaxies by stripping would lead to S0-like
galaxies, but not ellipticals, as stripping should not affect the
stellar content of the stripped galaxy. Besides, Burstein (1979)
observed that S0s seem to have thicker disks than spirals, which is
difficult to reconcile with the hypothesis that spiral galaxies
evolved into S0 by gas stripping.  The present model lies somewhat in
between the 'nature' and 'nurture' hypothesis. It is of interest to
test its predictions for the morphology-density relation. We undertake
this task in this section.

In order to evaluate the local morphology-density relation in our
model, we need a continuous relation between the collision factor
$\Delta$ and the density of the environment at the present
epoch. Scalings of $\Delta$ with density have been obtained in the two
limiting cases of the field (equation (\ref{eq20}), \S 3.2) and
cluster (equation (\ref{eq37}), \S 3.4) environments. However,
galaxies which are in a cluster now have not necessarily always been
in the same cluster environment.  In an $\Omega_0=1$ CDM universe,
clusters formed recently so that it seems fair to assume that energy
exchange among colliding galaxies in a cluster supplement the early
exchanges calculated for galaxies prior to cluster formation. In that
case, the collision rate $\Delta$ for a given galaxy can be taken as
\begin{equation}
\label{eq65}
\Delta=\Delta^{field}+\gamma \Delta^{cluster}
\end{equation}
where $\Delta^{field}$ is given by equation (\ref{eq20}) and
$\Delta^{cluster}$ by equation (\ref{eq23}) using the scaling with the
3-D galaxy density $n$ from equation (\ref{eq37}).  In (\ref{eq65}),
we have introduced the dimensionless parameter $\gamma$ to take into
account the fact that the thresholds $\Delta_{th}^{spi}$ and
$\Delta_{th}^{ell}$, that define the various morphological types
through the conditions (\ref{eq38})-(\ref{eq40}), may be different for
cluster and field galaxies.  According to (\ref{eq65}),
$\Delta^{field}$ dominates in the field, while the dominant
contribution to the collision factor is $\Delta^{cluster}$ in high
density environments. This defines a density $n$ at which
$\Delta^{cluster}$ starts to dominate over $\Delta^{field}$.
Observationally (Postman \& Geller 1984), the morphological
populations are rather constant up to $n\approx 1$ Mpc$^{-3}$
($h$=0.5). Accordingly, we fix the value of $\gamma$ so that
$\Delta^{cluster}$ starts to dominate over $\Delta^{field}$ at
$n\approx 1$ Mpc$^{-3}$. Numerically, using (\ref{eq20}) and
(\ref{eq37}), we find $\gamma \approx 10^{-3}$. We have considered the
case for an isothermal $\beta$-model for the cluster gas density with
$\beta=0.6$ (Jones \& Forman 1984). The cluster velocity dispersion
has been derived directly from the mass density profile using the
virial theorem.  Then, for $n\ll 1$ Mpc$^{-3}$, the proportions of the
various morphological types are the same as in the field, while for
$n\gg 1$ Mpc$^{-3}$, they are derived essentially from
$\Delta^{cluster}$.

We can now evaluate the present day morphology-density relation, using
equations (\ref{eq60})-(\ref{eq62}) where the redshift cuts $z_{ell}$
and $z_{spi}$ are now derived from (\ref{eq65}), with
$\Delta=\Delta_{th}^{ell}$ and $\Delta_{th}^{spi}$ respectively.
Again, the CDM spectrum (\ref{eq13c}) is used and is normalized to the
present abundance of X-ray clusters ($\sigma_8\approx 0.6$).  We use
the mass function (\ref{eq57})-(\ref{eq58}) with $\lambda=2$.  Figure
3 shows the morphology-density relation obtained from our modeling
(solid lines) compared to the observed relation (histograms; Postman
\& Geller 1984) at the present epoch. A typical 1-$\sigma$ error bar
is shown for reference in the top panel (from Postman \& Geller
1984). The model reproduces qualitatively the main features observed
in the evolution with density of the proportions of morphological
types: the spiral population decreases quite strongly in higher
density environments, while this decrease is compensated by a
corresponding increase of the elliptical and S0 populations.  The
agreement is actually rather good quantitatively also, at least for
low and intermediate densities. At high densities ($n\approx 1000$
Mpc$^{-3}$), the model overestimates slightly the fraction of spirals,
while the predicted fraction of ellipticals is somewhat lower than
observed.

Reproducing the observed morphology-density relation has required the
adjustment of 4 free parameters of our model: the two thresholds
$\Delta_{th}^{spi}$ and $\Delta_{th}^{ell}$; the parameter $\gamma$ in
(\ref{eq65}); and the parameter $\beta$ of the isothermal
$\beta$-model in (\ref{eq37}) .  The values of these parameters and
the method used to determine them is summarized in table 4.  The
predictions of the model are quite sensitive to $\beta$ and $\gamma$.
For example, if we use the assumption that mass traces light in
clusters, i.e. $\rho\sim n$, the model yields a decrease (an increase)
of the spiral (elliptical) population at high density {\it sharper}
than the ones observed.  If we vary the value of $\gamma$, the density
at which $\Delta^{cluster}$ starts to dominate over
$\Delta^{field}$ is different from the observed value $n\approx
1$ Mpc$^{-3}$.  It is however remarkable that the model simultaneously
reproduces the morphology-density for the three Hubble types
considered.  Once the model reproduces the observed morphology-density
relation today, quantitative predictions can then be made for the
evolution with redshift of the various fractions of morphological
types, in the field as well as in clusters.

\section{Evolution of galaxy populations with redshift}

\subsection{Redshift cuts}

\subsubsection{Field Galaxies}

Conditions (\ref{eq38})-(\ref{eq40}) on the collision factor $\Delta$
define the galaxy morphological types in our model. These conditions
have been shown to be equivalent to conditions on the redshifts of
formation of galaxies. The redshifts cuts on the redshift of
formation, $z_{ell}$ and $z_{spi}$, that delineate spirals from S0 and
from ellipticals are function of the redshift, since the collision
factor itself depends on the redshift.  Our model has been normalized
so as to reproduce the observed fractions of the different galaxy
populations today. It is a natural development of the model to predict
the evolution of these populations with redshift.

Equating $\Delta$ in equation (\ref{eq65}) to the thresholds
$\Delta_{th}^{ell}$ and $\Delta_{th}^{spi}$, whose values have been
determined previously, defines the two redshift cuts $z_{ell}(z)$ and
$z_{spi}(z)$ at any redshift $z$. On figure 4, $z_{ell}$ and $z_{spi}$
are plotted (thin solid lines) as a function of redshift, for field
galaxies. The field case is obtained by inserting the average galaxy
density of the universe into equation (\ref{eq65}) so that
$\Delta^{cluster}$ is always small compared to $\Delta^{field}$,
whatever the redshift. The functions $z_{ell}(z)$ and $z_{spi}(z)$
define three regions in the $(z,z_{nl})$ space corresponding to
specific ranges of values of $\Delta$, as indicated on the figure. At
a given redshift $z$, galaxies that went nonlinear at
$z_{nl}>z_{ell}(z)$ are identified as ellipticals, those with
$z_{nl}<z_{spi}(z)$ are spirals, and those whose redshift of formation
lies in between the two redshift cuts are S0s. The thick solid arrow
shows how a galaxy that formed at a given redshift $z_{nl}$ (here
$z_{nl}=3.1$) would be identified at different redshifts: down to
$z\simeq 2.7$, this galaxy is identified as a spiral, as long as
energy exchanges induced by tidal collisions are not high enough to
meet our morphology criterion for S0s. From $z\simeq 2.7$ to $z\simeq
1.8$, the galaxy is seen as an S0, before becoming an elliptical when
the cumulative effect of collisions since formation is important
enough to have induced substantial energy exchange.  In our scenario,
galaxies thus evolve along the Hubble sequence as time goes on.

\subsubsection{Cluster galaxies}

For cluster galaxies, the redshift cuts $z_{ell}(z)$ and $z_{spi}(z)$
are given by solving equation (\ref{eq65}) for
$\Delta=\Delta_{th}^{ell}$ and $\Delta_{th}^{spi}$ respectively, this
time with the average galaxy density of the universe replaced by the
typical local galaxy density $n$ of the cluster one wants to
model. Figure 5 shows $z_{ell}(z)$ for $n=500$ Mpc$^{-3}$ compared to
the field case.  For a given $n$, there is a redshift at which
$\Delta^{field}$ starts to dominate over $\Delta^{cluster}$. At high
redshift, $\Delta^{field}$ dominates, as expected since clusters
formed quite recently in a hierarchical universe.  At low redshift,
$\Delta^{cluster}$ dominates and $z_{ell}(z)$ is different from the
field values. In figure 5, $\Delta^{cluster}$ dominates over
$\Delta^{field}$ at $z\lta 1.5$. By construction, this redshift
reflects the formation epoch of the cluster and is higher for higher
values of $n$, since the cluster density reflects the density of the
universe at the epoch of its formation.

Another interesting feature in figure 5 is that the redshift cut
$z_{ell}$ is lower in a high density environment than in the field,
implying that cluster ellipticals form at a lower redshift than in the
field, as tidal collisions in clusters are more efficient than in the
field at converting disks into ellipticals.

\subsection{Redshift evolution}

At any redshift, the knowledge of the cuts $z_{ell}(z)$ and
$z_{spi}(z)$ allows one to count the number density of the different
galaxy populations.  As usual, we evaluate the fractions of the
various populations from equations (\ref{eq60})-(\ref{eq62}) with
$\eta(M,0,z_{ell})$ replaced, this time, by $\eta(M,z,z_{ell})$, etc.
All the results presented below assume the CDM spectrum (\ref{eq13c})
normalized to $\sigma_8=0.6$.

\subsubsection{Definition of irregular galaxies}

So far, we have considered only three morphological types.  However,
observations at high redshift show the existence of a large population
of so-called irregular galaxies (Brinchmann \etal 1997).  These
galaxies are morphologically perturbed, gas-rich, star forming
galaxies with blue colors.  There is a simple way to define
'irregular' galaxies in our model. Among all galaxies having
experienced substantial energy exchanges during tidal collisions
(essentially those that are ellipticals today), we can differentiate
between those galaxies that have relaxed by the time characterized by
redshift $z$ and those that have not.  The motivation for such a
distinction is that unrelaxed galaxies may exhibit the disturbed
morphology of observed irregulars, while relaxed galaxies are expected
to show a smoother stellar distribution, as in ellipticals.  Moreover,
tidal collisions are likely to induce star formation in the
interacting galaxies, and we can anticipate that the 'irregulars' of
our model should be star-forming galaxies and have blue colors, in
qualitative agreement with observations.  However, part of the
galactic gas is likely to be removed during collisions, and it is not
clear whether 'irregulars' in our model will have retained enough gas
to form stars.  Clearly, modeling of the effect of collisions on star
formation and the physics of the gas is required. Such models are
beyond the scope of the present paper and we shall restrict ourselves
to the present definition.

Our definition of 'irregular' galaxies thus amounts to introduction of
a new redshift cut, $z_{irr}$. It is fixed by requiring that the
elapsed time between the epoch of the last collision and the epoch
characterized by $z$ equals a typical time-scale for relaxation. Those
galaxies that were formed at redshift higher than $z_{irr}$ will have
relaxed by redshift $z$ and will be identified as ellipticals. In the
opposite case, these galaxies will be 'irregulars'.  In practice, we
assume that relaxation occurs on a time-scale $t_{relax}$ typically of
the order of a dynamical time-scale $t_{dyn}$, that is
\begin{equation}
\label{eq66}
t_{relax}\sim t_{dyn}\propto \sur{1}{\sqrt{G\rho}}\propto
(1+z_{nl})^{-3/2}
\end{equation} 
where $\rho$ is the density of the universe at the time of formation
$z_{nl}$. The epoch of the last collision is more problematic to
evaluate.  Actually, in our picture, they may not be any {\it epoch of
a last collision} as collisions continuously occur over the lifetime
of a galaxy. However, the bulk of collisions must occur at high
redshift when the density of the universe is high. The redshift
$z_{ell}$ (or a fraction of it) may be taken as the 'redshift of last
collision', though more minor collisions may occur at lower
redshift. There is some freedom in the choice of this epoch so that
the definition of irregulars in our scenario is somewhat loose.

\subsubsection{Evolution of field galaxies}

Figure 6 shows the evolution of the population content for field
galaxies.  The main observed trends in the evolution of the galactic
content of the universe are reproduced by our model: the fraction of
field spirals decreases with increasing redshift as collisions in the
past were more efficient at disrupting disk-forming galaxies.  They
make up 65\% of the galaxy population today and only about 15\% at
redshift 3.  The fraction of ellipticals rises smoothly from about
10\% today to about 20\% at redshift $\sim 0.5$, then starts decreasing
while the fraction of 'irregulars' takes over rapidly and starts to
dominate at redshift $z\sim 1$. We note that the substantial rise of
the 'irregular' fraction at redshift $z\sim 0.5$ is, at least
qualitatively, consistent with a recent analysis of the
peculiar/irregular populations of the CFRS and LDSS redshift surveys
(Brinchmann \etal 1997).

\subsubsection{Evolution of cluster galaxies}

Figures 7 and 8 show the evolution of the galaxy population with
increasing density of the environment.  We take $n=10$ Mpc$^{-3}$ in
figure 7 and $n=500$ Mpc$^{-3}$ in figure 8. At high redshift, where
$\Delta^{field}$ dominates over $\Delta^{cluster}$ in equation
(\ref{eq65}), the fractions of galaxy types are essentially the same
as in the field. Going towards lower redshifts, the fraction of spiral
galaxies increases less rapidly than in the field, peaks around $z\sim
0.3$ (figure 7) and steadily drops from 55\% to 50\% at the present
time.  This decrease is compensated by a corresponding increase of the
S0 and elliptical fractions, while the irregular population is quite
unaffected. Thus, the cluster core is depleted from its spiral
population as a direct consequence of the high density in the core,
while it is populated by S0 and ellipticals.  In figure 8, similar
behavior is noticed: the predicted fractions start to depart from the
field fractions at redshift $z\sim 1.5$, as expected from inspection
of figure 5, with the peak in the spiral fraction occurring at a
somewhat larger redshift than in figure 7.  The drop of the fraction
of spirals is more pronounced and goes from 45\% at redshift $\sim
0.6-0.7$ to 25\% at the present time.  It is temptative to associate
the decrease of the fraction of spirals from intermediate redshift to
the present with the well known Butcher-Oemler effect (Butcher \&
Oemler 1978; Dressler \etal 1994). Indeed, tidal collisions occurring
shortly after cluster formation might induce star formation in
spirals, that then would appear similar to the blue Butcher-Oemler
galaxies, while exhaustion and/or stripping of the gas content during
subsequent collisions would transform spirals into S0s and
ellipticals.  Clearly, inclusion of collision induced star formation
and gas physics in our model is needed for any detailed investigation
of its predictions with respect to the Butcher-Oemler effect.

\section{Conclusion}

We have argued that galaxy collisions play a fundamental role in
galaxy formation.  We have built a model in which galaxy morphological
types originate as a result of gravitational interactions with
surrounding galaxies.  Simple rules for energy exchange during
collisions have been proposed that allow us to discriminate between
different Hubble types: efficient collisions result in the disruption
of disks and substantial star formation, leading to the formation of
ellipticals; few or inefficient collisions in the past allowed a large
gas reservoir to survive and form disks. Quantitative analysis of
energy exchanges in the field and cluster environments have been
presented. These analyses are based on the simulation results of
Aguilar \& White (1985).

Assuming that galaxy formation proceeds in a $\Omega_0=1$ Cold Dark
Matter universe, our model both reproduces a number of observations
and makes various predictions, among which are the redshifts of
formation of the different Hubble types in the field. When the model
is normalized to the present day abundance of X-ray clusters, we find
that the amount of energy exchange needed to produce ellipticals
implies that they formed by $z\gta 2.5$ while spirals formed at $z\lta
1.5$.

The model also provides a natural basis for biasing between field
spirals and ellipticals without requiring an {\it ad-hoc}
identification of morphological types to peaks of different height in
the initial density field. Field elliptical galaxies are found to be
more biased with respect to mass than spiral galaxies by a factor
$\sim 1.5$.  They preferentially form out of $\sim 3\sigma$ peaks,
whereas spiral galaxies condense out of $\sim 2\sigma$ peaks.

Our formalism allows us to study galactic evolution in clusters. With
the same collision rules as in the field, the model satisfactorily
reproduces the morphology-density relation which spans a range of
densities from 10$^{-2}$ to 10$^4$ galaxies per Mpc$^3$.

Finally, the predictive power of the model is exploited to predict the
evolution of galaxy populations with redshift.  The predicted trends
are in good qualitative agreement with observations, both in the field
and in clusters. Our modeling of collisions in clusters naturally
gives rise to a Butcher-Oemler-like effect, observed in the predicted
spiral population. However, inclusion in our model of gas physics and
collision induced star formation is required for any detailed
quantitative comparison of the predictions of our model to
observations. Such a comparison will be undertaken in a forthcoming
paper.

\acknowledgments

We thank G. Mamon, A. Blanchard, P. Ferreira, B. Moore, S. Zepf and
R. Wyse for fruitful discussions or comments regarding the results
presented in this paper.  CB acknowledges financial support from the
Center for Particle Astrophysics and the Astronomy Department of the
University of California at Berkeley. This research has also been
supported in part by a grant from the NSF.

\newpage

%
%

\begin{center} 
Figures captions\\ 
\end{center} 
\vspace*{1cm} 

\figcaption[]{Average galaxy mass ${\bar M}$ (in $M_\odot$) as a function
of redshift, with and
without the cooling constraint (23). When cooling is included, ${\bar M}$
is relatively constant up to $z\simeq 4$.}

\figcaption[]{Number density of present day field galaxy populations as
a function of the redshift of formation of galaxies. This graph
shows the contribution of the three morphological types
defined in our model to the present population of galaxies. Galaxies
formed at redshift $z\ge z^{f}_{ell}$ become ellipticals by today.
Galaxies formed at $z^f_{spi}\le z \le z^f_{ell}$ are present day
S0s, while galaxies forming at $z\le z^f_{spi}$ are today's spirals.
The graph illustrates how morphological populations are defined and
counted
from a combined use of the mass function (66) and the predicted redshift
cuts $z^f_{ell}$ and $z^f_{spi}$}

\figcaption[]{Predicted local morphology-density relation (solid lines) 
compared
to the data (histograms) of Postman \& Geller (1984).
A typical one
standard deviation is shown on the top panel. The morphology-density
relation is computed in our model by assuming that energy exchange
in a cluster $\Delta^{cluster}$ 
supplement the early exchanges calculated for galaxies
prior to cluster formation $\Delta^{field}$.
Scaling of $\Delta^{cluster}$
with the local galaxy density $n$ is obtained by assuming that the cluster
density distribution traces the gas density distribution. 
An isothermal $\beta$-model is assumed
for the gas distribution.
}

\figcaption[]{Redshift cuts for the formation of
Hubble types, $z^f_{ell}$ and $z^f_{spi}$ (thin solid lines), are plotted 
as a function of
redshift in the field. $z^f_{ell}$ and $z^f_{spi}$ define three regions
in the ($z_{nl},z$) space, corresponding to three ranges of values of
the collision factor $\Delta$: ellipticals lie in the upper region,
Spirals in the lower region, while S0s lie in the intermediate region. 
The solid
thick arrow shows how the morphological type of a galaxy that was born
at redshift 3.1 evolves with redshift, under the effect
of collisions.}

\figcaption[]{The redshift cut $z^f_{ell}$ in the cluster environment 
(dashed line) is compared to the same cut in the field (solid line). 
At high redshift, the collision factor
(74) is given by the field limit $\Delta^{field}$, while at lower redshift
$\Delta^{cluster}$ starts to dominate over $\Delta^{field}$ and significant
discrepancy with the field appears. Ellipticals in a cluster environment
are able to form at lower redshift than in the field due to
the higher efficiency of collisions in clusters.}

\figcaption[]{Predicted fractions of galaxy populations in the field as
a function of redshift. A new morphological type has been
introduced and labeled as 'irregulars' (see text).}

\figcaption[]{Predicted fractions of galaxy populations for
$n=10$ Mpc$^{-3}$ as a function of redshift. 
At high redshift, predictions are the same as
in the field, whereas, at low redshift, significant differences
appear due to the different physics of collisions in denser environments.}

\figcaption[]{Same as figure 7 for $n=500$ Mpc$^{-3}$. Increasing $n$
accentuates the features of the evolution of the galaxy population
at low redshift.}
%
%
\begin{table}
\label{table1}
\caption[]{Name and meaning of parameters used}
\vspace*{0.5cm}
\begin{center}
\begin{tabular}{cc}
\hline \hline Name & Meaning\\ \hline \hline 
$R$ & Radius of the test galaxy\\ \hline 
$M$ & Mass of the test galaxy\\ \hline 
$V$ & Internal stellar velocity dispersion of the test galaxy\\ \hline 
$v$ & Relative velocity of encounters\\ \hline 
$\varepsilon$ & Mass fraction of the universe in galaxies\\ \hline 
${\bar \rho}$ & Average density of the universe\\ \hline 
$f$ & Normalization of the collision efficiency of AG85 function $f_E$\\
\hline 
${\dot \Delta}$ & Rate of change of binding energy due to collisions\\ 
\hline\hline
\end{tabular}
\end{center}
\end{table}
%
%
$\ \ $
\begin{table}
\label{table2}
\caption[]{Rate of change of binding energy}
\vspace*{0.5cm}
\begin{center}
\begin{tabular}{ccc}
\hline\hline Environment & Equation Used & Underlying Assumption\\
\hline\hline 
Cluster ($v\gg V$) & (\ref{eq10}) & Impulse Approximation\\
 & & Straight line trajectories\\ \hline 
Field ($v \sim V$) & (\ref{eq11}) & Gravitational focusing (\ref{eq9})\\ 
\hline\hline
\end{tabular}
\end{center}
\end{table}
%
$\ \ $
\begin{table}
\label{table3}
\caption[]{Numerical values of the model parameters}
\vspace*{0.5cm}
\begin{tabular}{ccc}
\hline\hline Parameter & Numerical Value & Units\\ \hline\hline
$\Omega_0$ & 1 & $\ldots$ \\ \hline
$H_0$ & 50 & km/s/Mpc\\ \hline
$M_*$ & $10^{12}$ & $M_\odot$\\ \hline
$R_*$ & 50 & kpc\\ \hline
$V_*$ & 200 & km/s\\ \hline
$z_*$ & 4.5 & $\ldots$ \\ \hline
$\varepsilon \rho_0/M_*$ & $1.75\times 10^{-3}$ & Mpc$^{-3}$\\ \hline
$v_0$ & 200 & km/s\\ \hline
$f$ & 30 & $\ldots$ \\ \hline
$\Delta_*$ & $1.75\times 10^{-4}$ & $\ldots$ \\ \hline
$\Delta_c$ & $3.5\times 10^{-4}$ & $\ldots$ \\ \hline\hline
\end{tabular}
\end{table}
$\ \ $
\begin{table}
\label{table4}
\caption[]{Tuned parameters for the morphology-density relation}
\vspace*{0.5cm}
\begin{tabular}{ccc}
\hline\hline Parameter & Numerical Value & Fixed By\\ \hline\hline
$\beta$ & 0.6 & Isothermal $\beta$-model\\ \hline
$\gamma$ & $10^{-3}$ & Galaxy populations at $n=1$ Mpc$^{-3}$\\ \hline
$\Delta_{th}^{spi}$ & 0.003 & Galaxy populations\\ \hline
$\Delta_{th}^{ell}$ & 0.01 & in the field today\\ 
\hline\hline
\end{tabular}
\end{table}

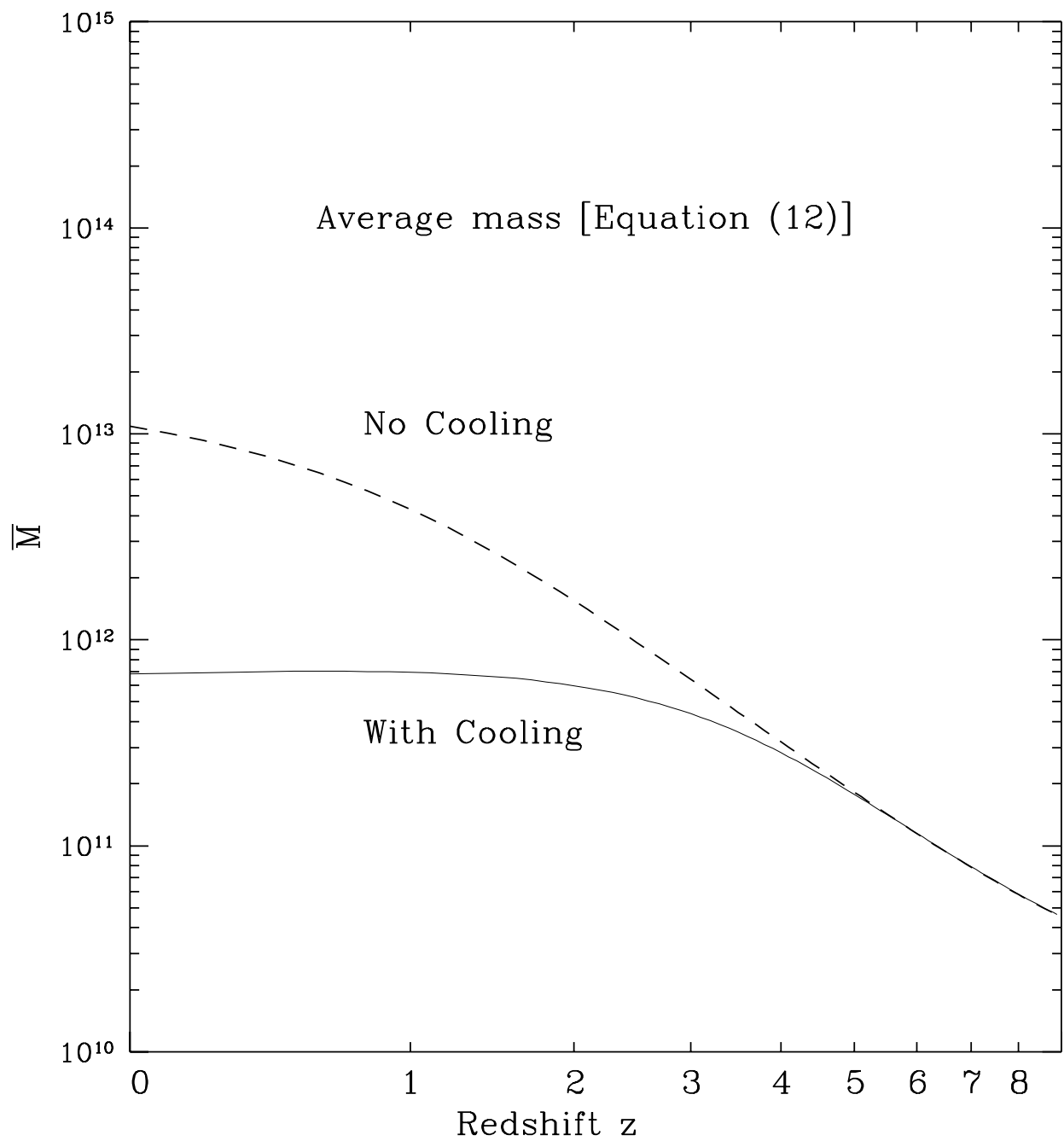
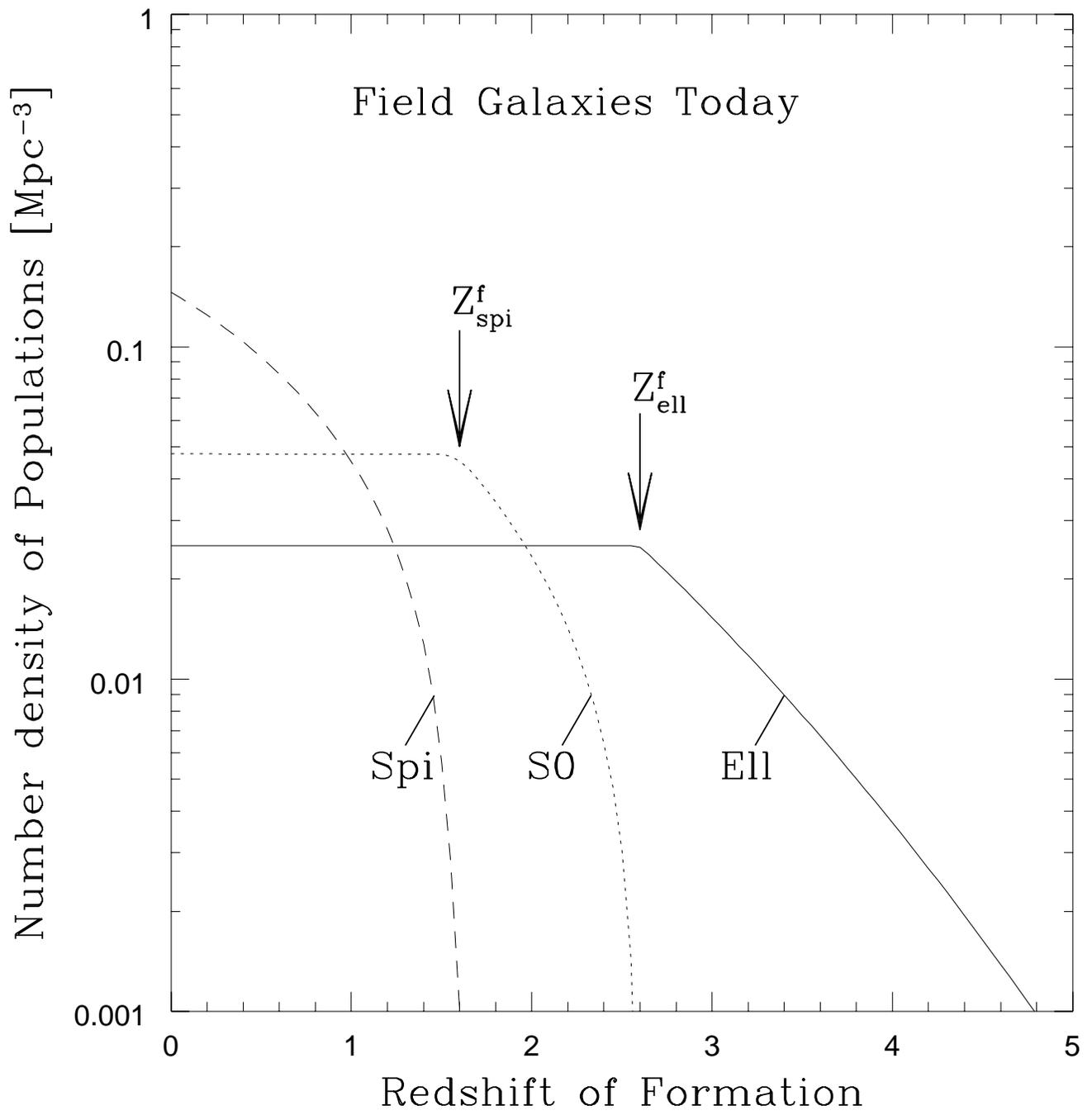
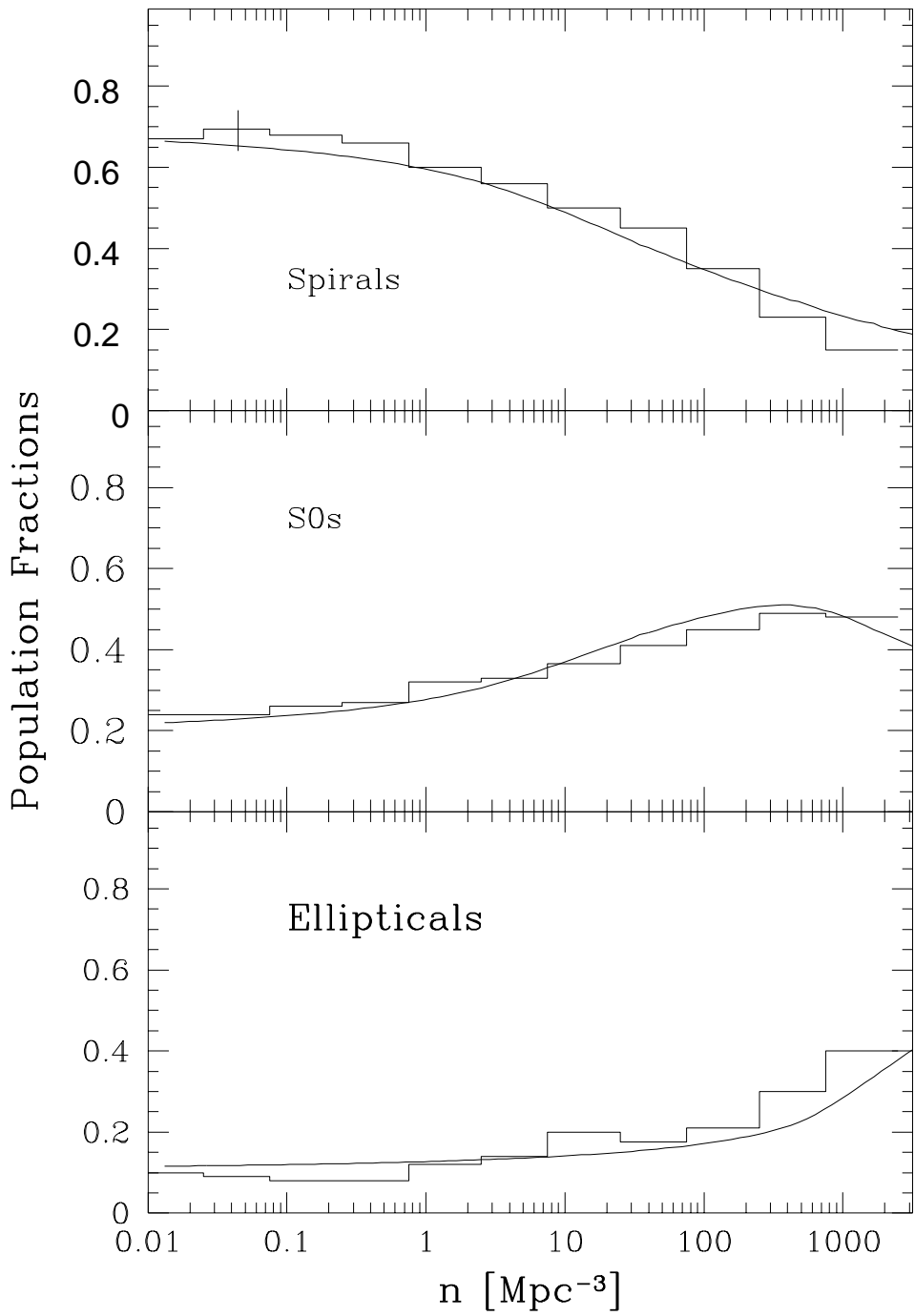
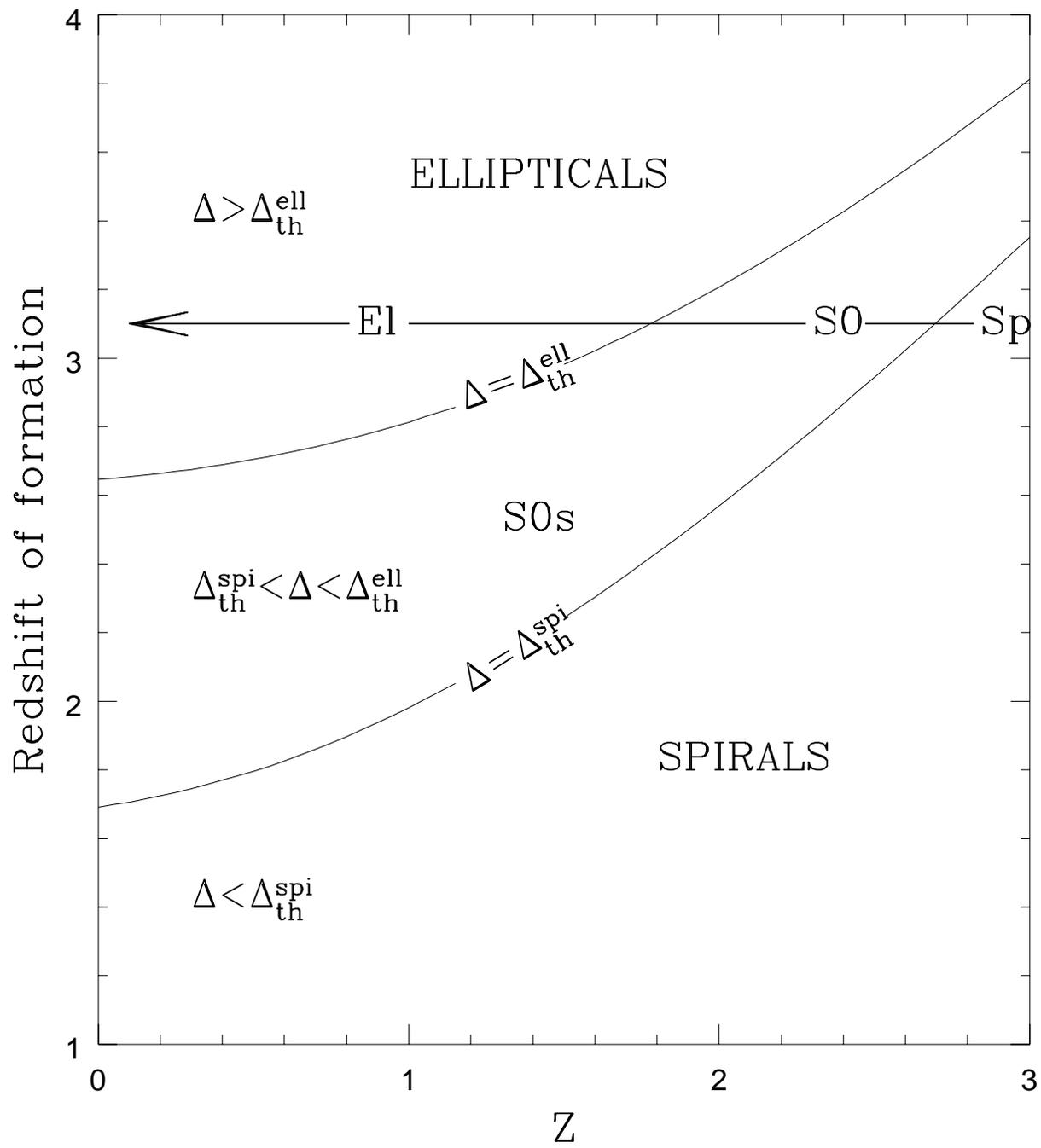
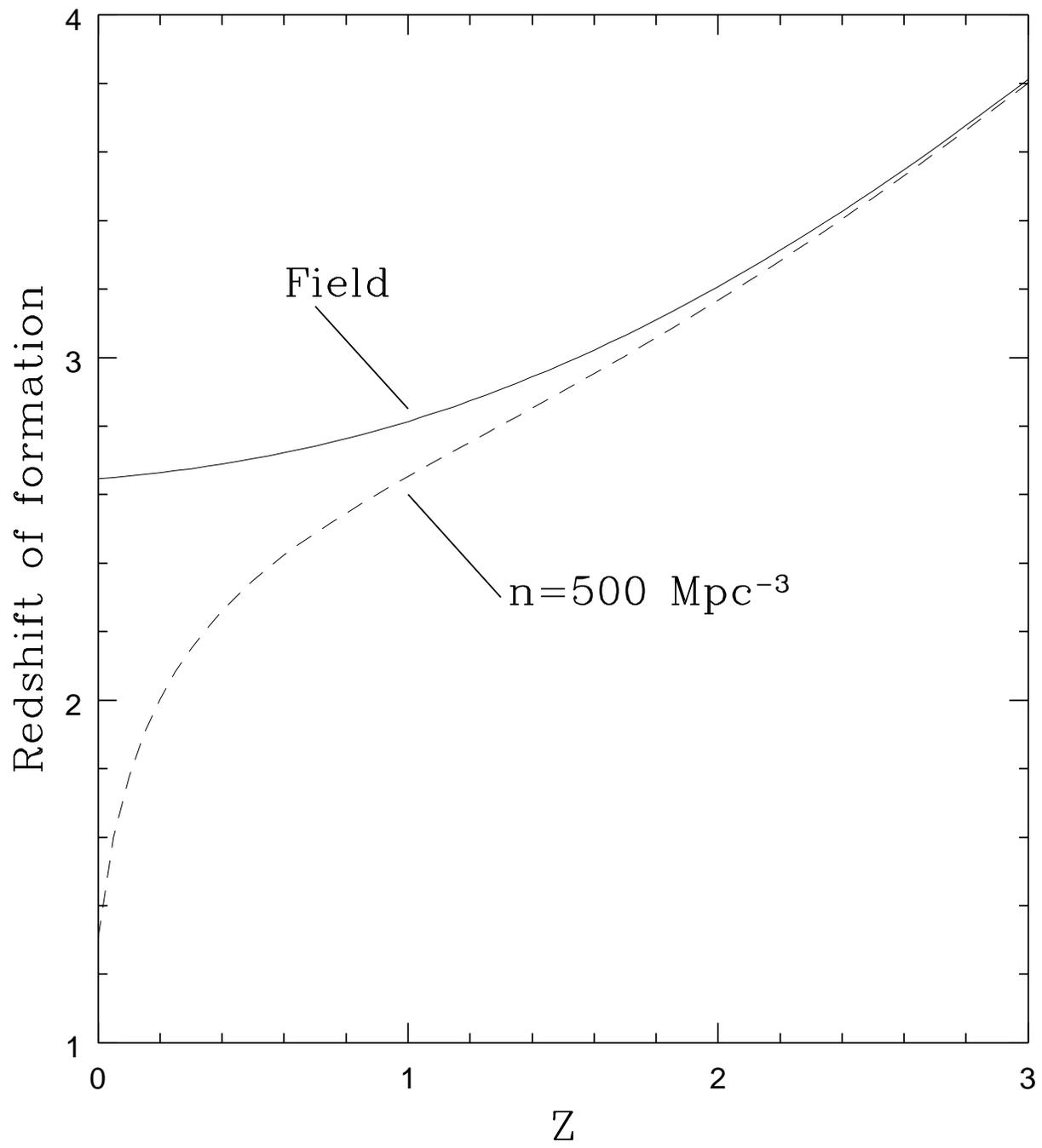
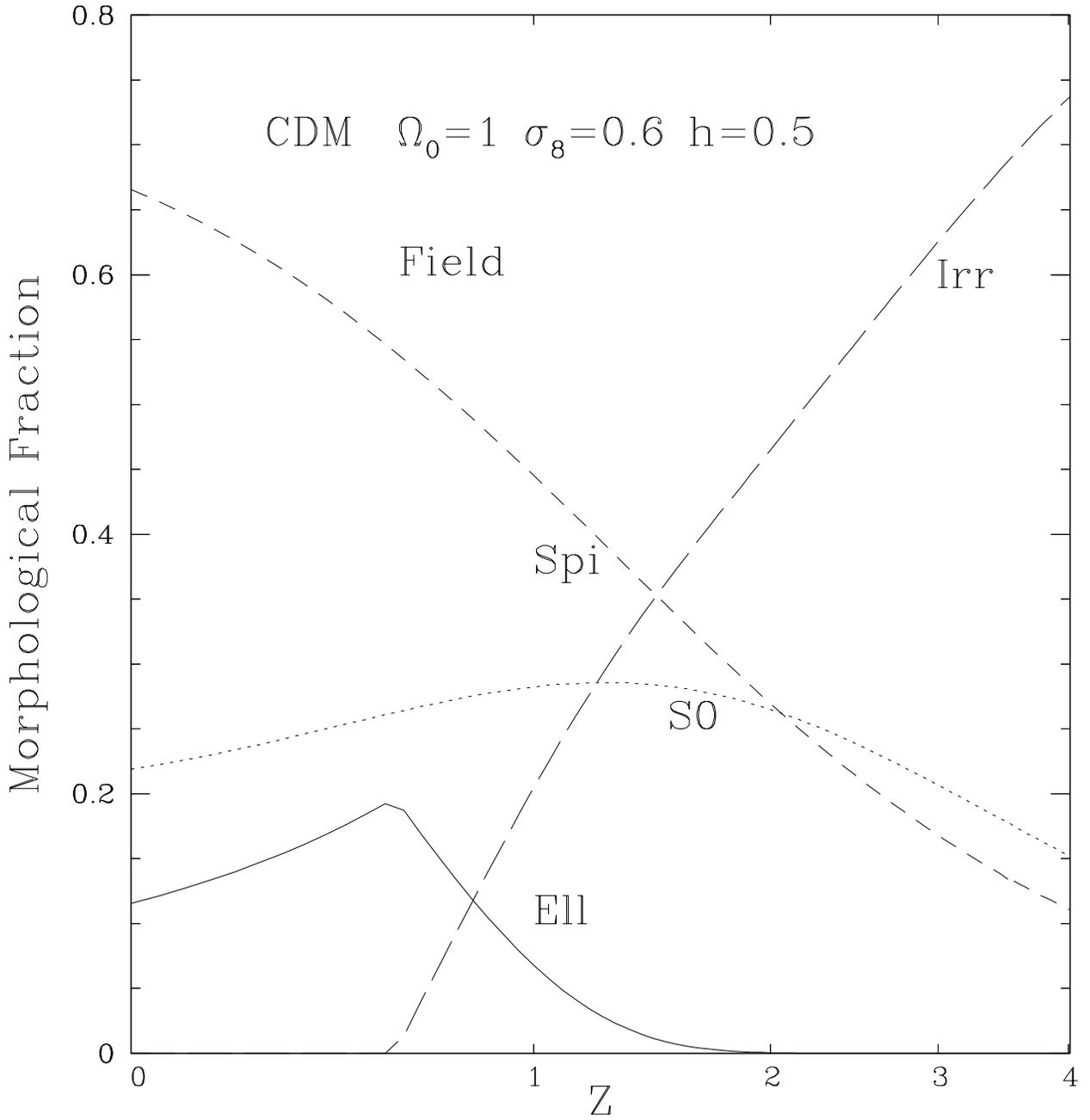
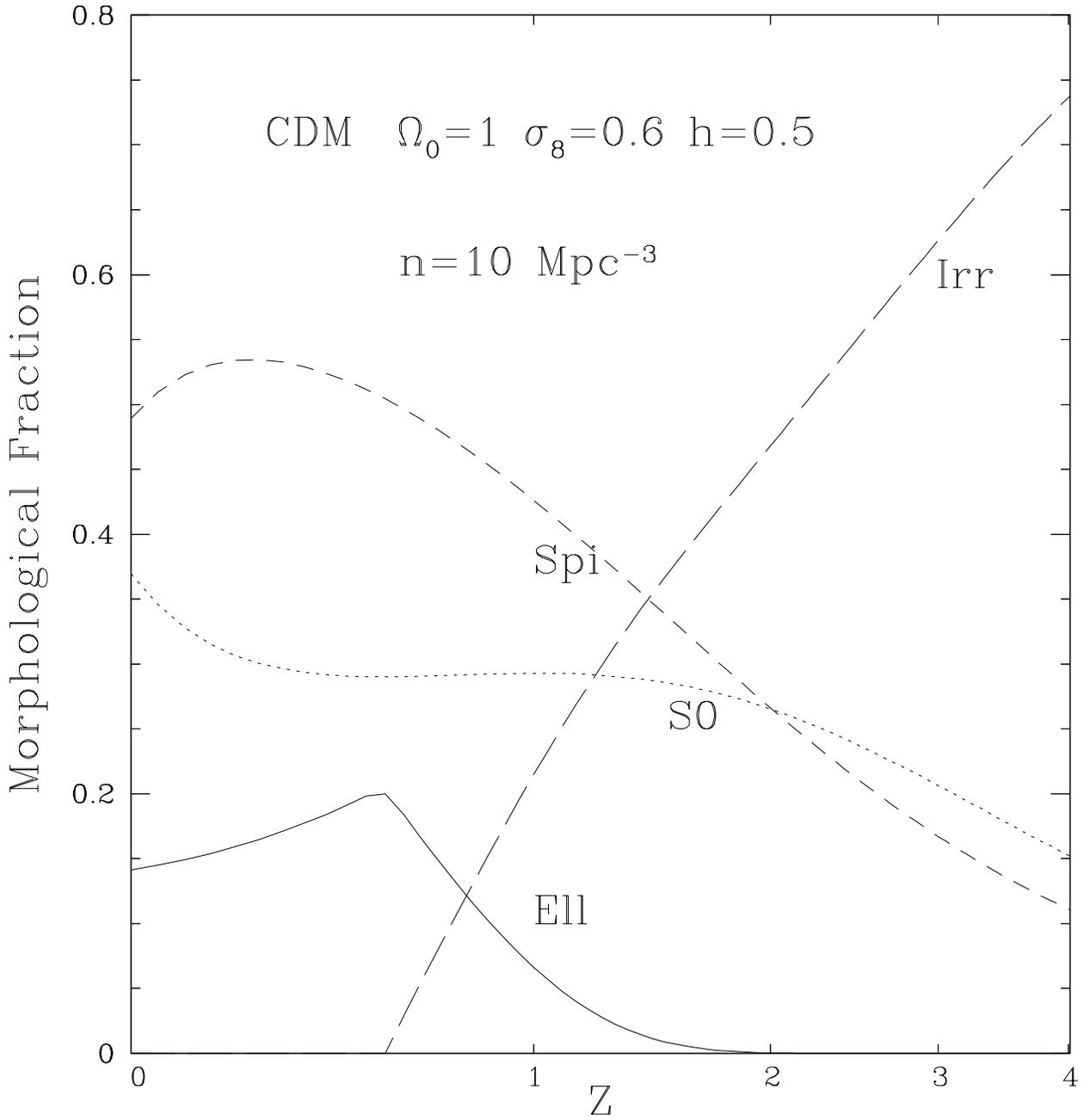
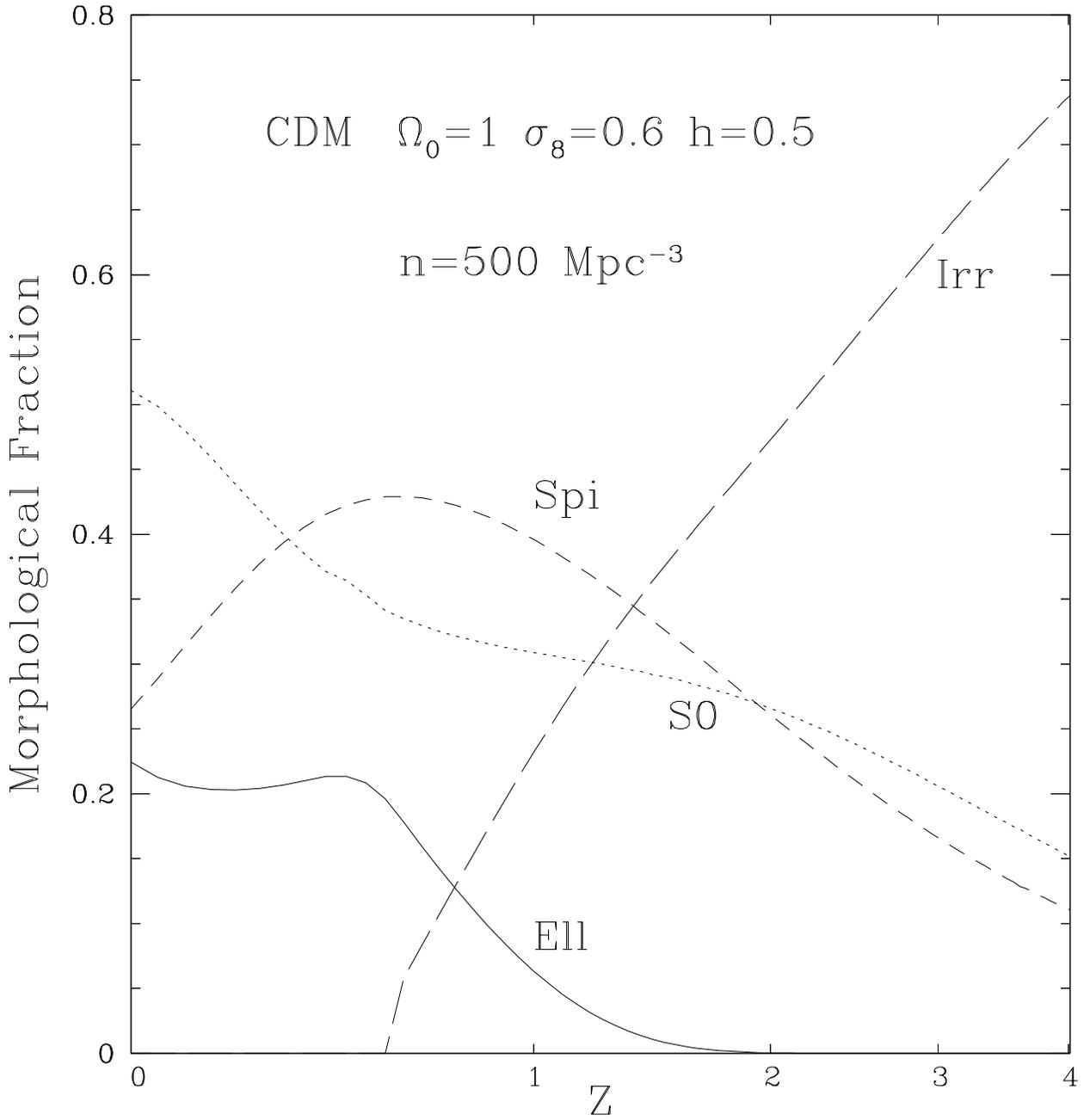
\end{document}